\documentclass{ar-1col-S2O}
 
\usepackage[sort&compress,numbers]{natbib}
\usepackage{url}
\usepackage{xcolor}
\usepackage{soul} 
\usepackage[normalem]{ulem}
\usepackage{graphicx}
\usepackage{bm}
\usepackage{float}
\usepackage{longtable}
\usepackage{wrapfig}
\usepackage[colorlinks=true,allcolors=red]{hyperref}

\usepackage[colorinlistoftodos,prependcaption]{todonotes}

\jname{Xxxx. Xxx. Xxx. Xxx.}
\jvol{AA}
\jyear{YYYY}
\doi{10.1146/((please add article doi))}

\begin{document}

\markboth{V. S. Sivasankar and R. N. Zia}{The matter/life nexus in biological cells}

\title{The matter/life nexus in biological cells}

\author{Vishal S. Sivasankar and Roseanna N. Zia$^\dagger$
\affil{Department of Mechanical and Aerospace Engineering, University of Missouri, Columbia, United States, MO 65211; $^\dagger$Correspondendence: rzia@missouri.edu}
}

\begin{abstract}
    The search for what differentiates inanimate matter from living things began in antiquity as a search for a ``fundamental life force” embedded deep within living things --- a special material unit owned only by life --- later transforming to more circumspect search for unique gains in function that transform nonliving matter to that which can reproduce, adapt, and survive.  Aristotelian thinking about the matter/life distinction and Vitalistic philosophy’s “vital force” persisted well into the Scientific Revolution, only to be debunked by Pasteur and Brown in the 19th century. Acceptance of the atomic reality and understanding of the uniqueness of life’s heredity, evolution, and reproduction led to formation of the Central Dogma. With startling speed, technological development then gave rise to structural biology, systems biology, and synthetic biology --- and a search to replicate and synthesize that ``gain in function” that transforms matter to life. Yet one still cannot build a living cell \textit{de novo} from its atomic and molecular constituents, and ``that which I cannot create, I do not understand”. In the last two decades, new recognition of old ideas --- spatial organization and compartmentalization --- have renewed focus on Brownian and flow physics. In this Review, we explore how experimental and computational advances in the last decade have embraced the deep coupling between physics and cellular biochemistry to shed light on the matter/life nexus. Whole-cell modeling and synthesis are offering promising new insights that may shed light on this nexus in the cell’s watery, crowded milieu.

\end{abstract}

\maketitle

\tableofcontents


\section{Introduction}
Engineers and scientists studying biological cells and flows aim to harness, optimize, and mimic cellular functions. Yet cells’ remarkable machinery often defies copying or even modification – especially the elements unique to cellular functions – inevitably inspiring the question: where or how does matter become life? Even a partial answer requires recognition of what differentiates matter from life, a question that has driven millennia of technological and theoretical development. This arc of inquiry has evolved from philosophy to characterization and now to synthesis — the idea that ``if I can build it, I understand it." To frame our perspective of the last few decades’ advances in this area, we first briefly review the wider history of how scientists have differentiated matter from life.

Early philosophers defined life as `agency.' The Vitalistic thinkers of ancient Greece linked autonomous motion to agency and life through Aristotle's \textit{entelechy}, a mysterious ``life force" \citep{aristotle365anima, aristotle1986anima} thought to drive autonomous motion. This link was also popular among Materialists like Epicurus (300 B.C.E.) and Lucretius (50 B.C.E.), who connected life to matter's motion through Epicurean physics \citep{mayr1982growth}, which paved the way for modern molecular and evolutionary models of life. Early questions about `what is life' thus centered on how to explain agency and autonomous motion.

\begin{marginnote}[]
	\entry{agency}{the capacity to act with autonomy}
	\entry{teleology}{``The use of ultimate purpose or design as a means of explaining phenomena" \cite{AmHtgDict}, {\em e.g.}, flowers' structures develop with intent to attract pollinators.}
\end{marginnote}

But during the Scientific Revolution, natural philosophers (scientists) \citep{schuster2006scientific} replaced Aristotelian philosophy with systematic empirical observation, experimentation, and inductive reasoning. This approach produced predictive theories that mechanistically explained phenomena. It also demanded new technology, such as the microscope, which pushed the natural sciences toward the microscopic — and the idea of fundamental building blocks of life. As an example, Leeuwenhoek's microscopic observations of ``animalcules" – tiny organisms moving unassisted – reinforced the contemporaneous idea that motion is a defining feature of life \citep{dobell1932antony}. Thus, questions about `what is life' continued to center on autonomous motion, but inquiry moved to a smaller scale in search of its underlying origin.

\begin{marginnote}[]
	\entry{Aristotelian thinking}{the idea that understanding living physics requires understanding the purpose of the physics}
	\entry{Scientific Revolution}{1543-1687. A period of significant developments in mathematics, physics, astronomy, biology and chemistry that transformed understanding of nature.}
\end{marginnote}

In the early 1800s, Robert Brown discovered autonomous motion in living pollen grains, but his rigorous methods revealed that the curious autonomous motion did not require life at all, only that the particles under observation be small \citep{brown1828xxvii}. Eighty years later Einstein proved mathematically and Perrin demonstrated experimentally that this Brownian motion reveals matter's atomic nature rather than the fundamental force of life \citep{einstein1906theory, smoluchowski1906kinetic, perrin1909movement}. 

Unknowingly, Brown had identified thermal motion and attributed it to any matter, not just living matter. Many tools used to study Brownian motion in cells and other biological flows trace back to these first microscopy experiments and the Stokes-Einstein relation. Haw's \textit{Middle World: The restless heart of matter and life} describes this remarkable time in a masterful, inspiring, and scholarly story \cite{haw2006middle}. In his effort to discover where matter becomes life, Brown’s observations instead ended the long-standing notion that autonomous motion is synonymous with life.

\begin{marginnote}[]
	\entry{Brownian motion}{random walk of microscopic particles through liquid, induced by individual, randomly directed impacts of solvent molecules.}
	\entry{Stokes-Einstein relation}{a particle's diffusion $D$ dissipates thermal energy $kT$ viscously back into a solvent of viscosity $\eta$, as $D=kT/6\pi\eta a$, for particle size $a$.}
\end{marginnote}

Recognition that \textbf{only living matter can reproduce} emerged in the mid-1800s, replacing autonomous motion as the differentiator of matter from life. Schwann and Shleiden \citep{schleiden1838beitrage, schwann1839mikroskopische} led in this area, postulating that \textbf{cells are the fundamental units of life} responsible for reproduction. Their ideas trace back to Leeuwenhoek's identification of spermatozoa (1676) \citep{leewenhoeck1977observation}, Bauer's orchid cells (1802) \citep{bauer1838illustrations} and Brown's nucleus (1831) \citep{brown1833observations}. This definition of life as that which can reproduce and grow became the basis of modern biology.

The search for how cells reproduce transformed the question ``what is life" into a series of questions that became fields of study: ``what are cells made of?" (omics), ``how do cells survive?" (evolution), and ``how do cells propagate?" (heredity and growth). Theories of inheritance, genetic variation, natural selection, and evolution emerged and formed the basis of modern evolutionary biology \citep{mendel1996experiments,huxley1942evolution,darwin2016origin,de1910intracellular, muller2010cell}. Meanwhile, detailed microscopy studies carefully cataloged cellular constituents, including the first observation of a nucleus \citep{brown1833observations} and chromosomes \citep{vines1880works}. In 1869, Miescher isolated DNA and its associated proteins from leukocyte and \textbf{hypothesized that DNA is the material basis of heredity} \citep{miescher1871ueber}. Yet, explicit description of the physics of this material took time to emerge.

\begin{marginnote}[]
	\entry{RNA}{ribonucleic acid.}
	\entry{mRNA}{messenger RNA}
	\entry{DNA}{de-oxyribonucleic acid}
\end{marginnote}

\begin{marginnote}[]
	\entry{Rosalind Franklin}{British chemist whose X-ray diffraction images of DNA enabled characterization of its double-helix structure.}
\end{marginnote}
In 1958, inquiry around omics, evolution, and heredity converged to form the Central Dogma, a comprehensive model of cellular life \citep{crick1957nucleic}. RNA had been discovered \citep{altmann1889ueber} and the nucleic acid composition of ribose sugar, phosphate and  nitrogenous bases was established \citep{levene1931nucleic}.
\textbf{Structure-function relationships} emerged: using DNA's chemical composition and Franklin's X-ray crystallography \citep{franklin1953molecular}, Watson and Crick determined the double-helix structure of DNA \citep{watson1953structure}.This structure was then used to deduce DNA’s function: it stores and reliably transfers genetic information, providing \textbf{the physico-chemical basis of heredity}. 
Crick further inferred that RNAs serve as intermediaries between DNA and proteins, and proposed a genetic information flow inside cells --- the \textbf{Central Dogma hypothesis} \citep{crick1957nucleic}, \textbf{recasting our understanding of cellular life into an information handling process}. Although DNA's physical structure sets its function, information processing was the startling new discovery that dominated the study of life for decades. 

The Central Dogma advanced the understanding of where matter becomes life by defining the process by which life creates itself. But a piece was missing: to survive a changing environment, cells must gain functions, suggesting that the matter-to-life leap requires survival and the ability to adapt.

In the 1960s scientists discovered that \textbf{DNA physically restructures to enable adaptation} via processes outside the Central Dogma. Monod and subsequent workers \citep{jacob1961genetic,roeder1969multiple, ullmann1971cyclic} found that genes produce environmentally-responsive transcription factors – like the catabolite activator protein (CAP) and histone-like proteins – that physically condense, coil, and bend gene segments to expose genes and regulate transcription. Mutation bias is yet another example of such ``living physics" \citep{boulikas1992evolutionary,tomkova2018dna, duan2018reduced} not captured by the Central Dogma.

Overall, the search for where matter becomes life drove a march down to the atomistic length scale, which culminated in a startlingly clear mechanistic model for life’s general dogma. But declaring the Central Dogma as the entire answer to how matter becomes life seems reductive. Even understanding the atomistic elements of DNA and RNA, one cannot `boot up' new life by placing them in a test tube alongside all the metabolites and growth media needed and shaking it up. Without all the other parts of the cell, the Central Dogma does not translate to life. Thus, in the 1990s, the inquiry about where matter becomes life moved back upwards to whole-cell function.

This is where we pick up the story. In this review, we cast a wide net into cellular biophysics approaches that shed light on how matter becomes life in biological cells. We focus on essential thematic areas: structural biology and systems biology; genetic engineering and synthetic cells; cytoplasm organization and compartmentalization; the role of fluid mechanics in cellular processes; and finally, whole-cell models. Each section touches on major milestones and landmark achievements via experimental, statistical, and computational approaches.

\section{The building blocks: structure, function, and the Central Dogma}
Early 20$^{th}$ century studies of DNA, RNA, ribosomes, and proteins suggested that macromolecules' physical structure plays a key role in the transformation of matter to life. The discovery of DNA’s structure via X-ray crystallography confirmed this idea and gave rise
to structural biology, which demonstrates how molecular-scale structures (e.g. helices, coils, folding) drive biological functions (e.g., transcription, translation, replication, signaling, chaperoning). Studying structure/function relationships drove development of many new experimental techniques, including X-ray crystallography, nuclear magnetic resonance (NMR), cryo-electron microscopy (cryo-EM), small angle X-ray scattering (SAXS), small angle neutron scattering (SANS), fluorescence microscopy techniques, single molecule tracking (SMT) and other super-microscopy techniques \citep{curry2015structural,sahl2017fluorescence,schermelleh2019super,bacic2020recent}. Discoveries from these techniques are cataloged in the Protein Data Bank (PDB), an open-source repository for over 200,000 proteins, ribosomes, nucleoprotein complexes \citep{burley2023rcsb}, and other structures that shed light on cellular-level disease \citep{morimoto1971molecular,  wang2016exploring, renaud2020structural}, ribogenesis, enzymatic reactions \citep{sacquin2016bridging}, and more.

\begin{marginnote}[]
	\entry{structural biology}{connecting molecular structure and dynamics of macromolecules to their biological functions.}
	\entry{X-ray crystallography}{using X-ray diffraction to reveal 3D arrangement of electrons, atoms, and bonds within molecules}
\end{marginnote}

DNA's first structure/function relationship was made possible by Franklin's careful X-ray crystallographs \citep{franklin1953molecular}, which upended contemporaneous speculation of a triple helix structure \citep{pauling1953proposed} in favor of a double helix. Once obtained and analyzed by Watson and Crick \citep{watson1953structure}, they speculated that DNA's double stranded, stacked chain structure constrained specific complementary base-pairs into a template by which DNA replicates \citep{crick1954complementary}. Evidently the matter-to-life transition requires stable, ordered structure within DNA. Crick's subsequent insight that the nucleotide sequence stores and transfers genetic information led to his Central Dogma hypothesis \citep{crick1958protein}, where life is understood as a flow of information: transcription (DNA makes RNA), translation (RNA makes proteins), and replication (DNA copies itself). For decades after, a DNA-centric view of life focused on sequence and information processing \citep{thieffry1998forty, hewitt2020negative}.

\begin{marginnote}[]
	\entry{Central Dogma}{DNA makes RNA via transcription; RNA makes proteins via translation; DNA copies itself via replication}
\end{marginnote}

\begin{textbox}[t]\section{PROTEIN COMPLEXATION: A FUNDAMENTAL GAIN IN FUNCTION}
	Pauling and Corey's measurements based on interatomic distances and bond angles led them to hypothesize that atomic chemistry such as hydrogen bonding drives proteins to form secondary structures, $\alpha$ helices and $\beta$ sheets \citep{pauling1951structure}. They parlayed these ideas to speculate formation of coils and sheets into tertiary structures, later confirmed by X-ray crystallography \citep{hodgkin1951crystallography}.
\end{textbox}

But subsequent discoveries revealed that proteins and other biomolecules undergo conformational changes that enable gain or loss of life-critical functions \citep{whitford2013proteins}. Pauling and others recognized that organic molecules exhibit far more complex structures than inorganic macromolecules \citep{hargittai2015linus}, leading to the landmark concept of protein folding. \textbf{Complexation of proteins via ``folding" is as important to function as amino acid sequence}, creating structured and disordered regions that are vital to biological function \citep{whitford2015protein}. For example, small structural changes undermine hemoglobin’s oxygen-binding ability and disable antibodies' binding to pathogens. Predicting three-dimensional folded structure directly from the amino acid sequence has been identified as one of the top 125 outstanding scientific problems of our time \citep{american2005so}. Protein \textit{mis}folding can produce aggregates that underlie disease like sickle cell anemia, amyloidosis, type-2 diabetes, Alzheimer's, Parkinson's, Huntington's, and ALS \citep{chiti2006protein,valastyan2014mechanisms}. Discoveries in this area pushed ``structural genomics" to the forefront of drug discovery \citep{morimoto1971molecular, wang2016exploring, renaud2020structural} and drove development of algorithms to predict complex tertiary structure from data in the Protein Data Base, another landmark development  \citep{berman2000protein, berman2003announcing, wwpdb2019protein}. Such algorithms include template-matching methods like homology modeling and threading \citep{qu2009guide, fiser2010template} as well as template-free modeling such as \textit{ab initio} energy minimization techniques like Baker's Rosetta \citep{bonneau2001rosetta, rohl2004protein}. PDB data are now also exploited by machine learning algorithms like AlphaFold that use multiple sequence alignment (MSA) techniques to predict many new protein structures \citep{jumper2021highly, evans2021protein, varadi2022alphafold, varadi2024alphafold}.

\begin{marginnote}[]
	\entry{structural genomics}{aims to determine 3D structure of every protein encoded by a genome, {\em e.g.}, for structure-based drug discovery}
	\entry{tertiary structure} {a protein complexate formed through folding of secondary protein structures ($\alpha$ helices, $\beta$ sheets)}
	\entry{protein structural prediction}{aims to predict 3D folded structure using only amino acid sequence}
\end{marginnote}

A third landmark in protein study is the discovery of intrinsically disordered domains
(IDDs), which are vital to healthy physiology and implicated in pathology \citep{deiana2019intrinsically}. Hydrophobic
cores that drive formation of globular structures are missing in these IDDs, producing energetically frustrated states with local minima that thwart stably-folded conformations and cause structural ambiguity \citep{rezaei2012intrinsically}. Approaches such as refined force fields have enhanced prediction of intrinsically disordered protein ensembles \citep{rauscher2015structural}. Their structural flexibility enables IDPs to bind without specificity, which underlies their propensity to aggregate and phase separate \citep{strodel2021energy}. Recent study of prion-like proteins shows that ordered and disordered regions responsible for biological function also cooperate to form stable, heritable aggregates \citep{yang2024in-prep}. Prions thus reproduce themselves outside the Central Dogma, blurring the line between biological material and life.

\begin{marginnote}[]
	\entry{intrinsically disordered proteins}{continuously fluctuating conformations lack stably-ordered domains.}
\end{marginnote}

Allosteric regulation, key in a cell’s response to environmental stress \citep{nussinov2012allosteric, macek2019protein}, also demonstrates matter-to-life transformation outside the Central Dogma. It occurs via chemical post-translational modifications that allow proteins to fold differently than the encoded amino acid sequence dictates. Gr{\"a}ter and co-workers have shown how such post-translational modifications alter proteins’ structure and functions \citep{jin2021multisite}. 

Many of these essential post-translational, stress-responsive, and complexation behaviors require cooperative coordination between proteins and other cytoplasmic molecules. But despite the uniqueness of proteins’ individual physics compared to non-biological macromolecules, a protein’s gains in function do not instantiate life. Somewhere in the interconnectedness of protein functions, life emerges from matter. Systems biology focuses on this interconnection.

\section{Systems biology: networks that weave life together inside the watery cell}
\label{sec:systemsbio}
Systems biology aims to represent a whole cell's functionality --- including genome, proteome, metabolome, and other atomic and ion functions --- as chemical reactions via a master network of interconnected reaction pathways \citep{kitano2002systems,aggarwal2003functional, costa2008complex}. The pathways are built from experimental and \textit{ab initio} measurements of individual reaction rates combined with whole-cell functional measurements during systematic gene knockouts, effectively mapping tens of thousands of reactions. This Edisonian effort maps each constituent molecule onto a predictive network as environmental conditions change. This whole-cell approach has yielded remarkable insights about enzymatic, signaling, metabolic, and anabolic reaction pathways that ultimately translate genotype to phenotype. The resulting ability to predict adaptation is one of the superpowers of this approach. 

\begin{figure}
\centering
\includegraphics[width=0.9\linewidth]{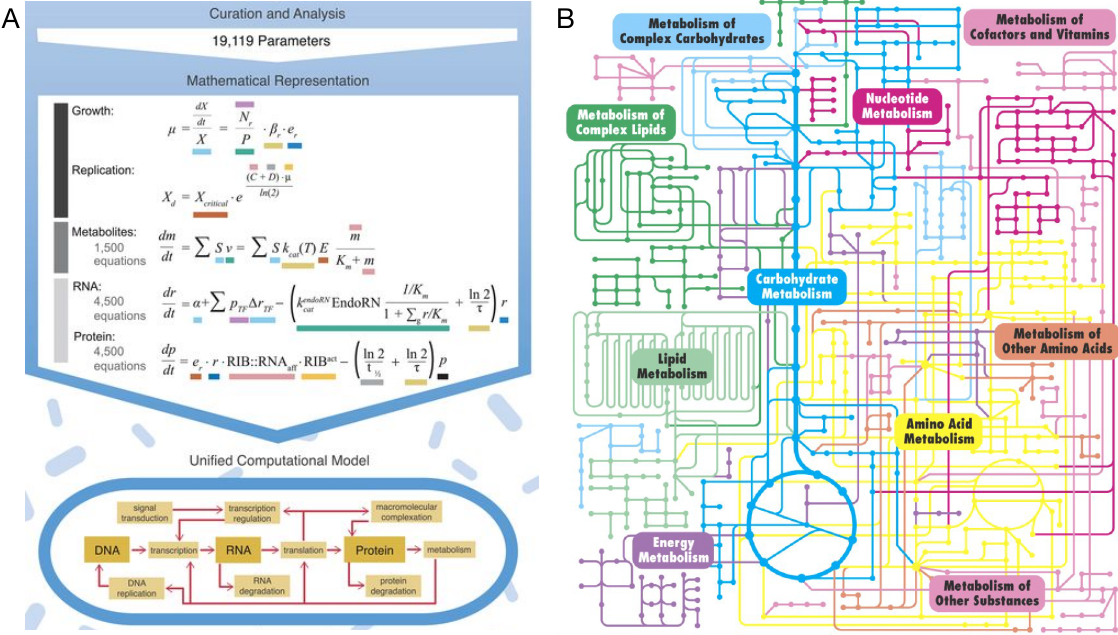}
\vspace{-1 mm}
\caption{\footnotesize   (A) Mathematical description of the kinetics-based whole-cell model of \textit{E. coli} [adapted with permission from \citep{macklin2020simultaneous}]. (B) Illustration of a metabolic pathway [permission requested from \citep{hillis2012principles}].}
\label{fig:fig3} 
\vspace{-2 mm}
\end{figure}

\begin{textbox}[t]\section{Landmark achievements of systems biology}
	Systems biology's landmark achievements include conceptualization of genetic circuits \citep{mcadams2000gene,walleczek2006self,brophy2014principles}, which advanced understanding of gene expression regulation \citep{mcadams1998simulation}, metabolism \citep{kim2018molecular}, translation \citep{zhao2014mrna}, cell-fate decisions \citep{guillemin2020noise}, and phenotype switching \citep{thomas2014phenotypic}, and the elucidation of metabolic networks for reactions and signaling pathways \citep{cascante2008metabolomics}. Together, these paved the way for genome-wide, integrative systems biology models such as Endy's TABASCO with base-pair level resolution \citep{kosuri2007tabasco}. Covert transformed these ideas into whole-cell, full-cycle kinetic models of \textit{Mycoplasma genitalium} and {\textit{E. coli}} [Fig. \ref{fig:fig3}A]  \citep{karr2012whole,maritan2022building, macklin2020simultaneous}. These interconnected networks are represented in circuit-like diagrams [Fig. \ref{fig:fig3}B].  More recently, data-driven machine learning models have filled in numerous network gaps \citep{shah2021review,presnell2019systems}. 
\end{textbox}

\begin{marginnote}[]
	\entry{systems biology}{study of interactions between interconnected network of cellular constituents to predict whole-cell function}
	\entry{genotype to phenotype}{how biochemistry and genetics lead to a range of survival and growth outcomes based on environmental conditions}
\end{marginnote}

Yet, these couplings' deep complexity obscures exactly where matter transforms to life.  In a sense, the completeness of traditional whole-cell modeling provides too much information. In these vast networks, in which process -- at what scale -- in which pathway -- does matter become life? Synthetic biologists might answer this question by claiming that if one can build life, one understands life.

\section{Synthetic cells: build life to understand it}
\label{sec:SynthCells}
\noindent Synthetic biology formalized modern systematic exploration of the elusive boundary between matter and life \citep{deplazes2009synthetic, ginsberg2014synthetic}. Synthetic biology grew out of desires in the 1950s to discover ``the smallest autonomous self-replicating entity'', as put by Harold Morowitz \citep{morowitz1984completeness}. The focus quickly ``shifted from the smallest organism to the smallest genome.'' \citep{morowitz1984completeness}. The ability to sequence genomes ---  along with tools to manipulate genes --- made this quest a reality
\citep{old1981principles}. These techniques followed from the discovery of restriction enzymes and development of CRISPR techniques \citep{di2019restriction} as well as advancements in DNA cloning and PCR technologies \citep{brown2020gene}. There are numerous excellent reviews of the field \citep{benner2005synthetic, hanczyc2020engineering, bashor2018understanding, ren2020recent, weninger2016key}. These techniques are all brought together in bottom-up and top-down techniques for building synthetic cells, with frontiers being advanced by open-collaboration research communities like \textit{Build A Cell} (https://www.buildacell.org) led by Adamala and co-workers.
\begin{marginnote}[]
	\entry{restriction enzymes}{self-guiding proteins that cleave DNA at a specific site to form DNA fragments}
	\entry{CRISPR}{technique to edit genes by cleaving DNA at precise sites via Cas enzymes guided by guide RNA.}
\end{marginnote}

\begin{figure}
\centering
\includegraphics[width=0.85\linewidth]{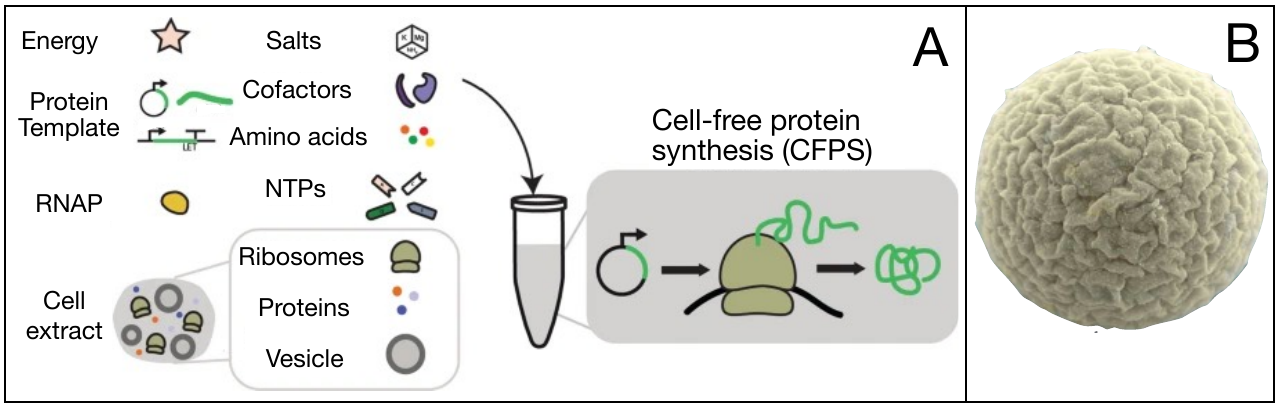}
\vspace{-2 mm}
\caption{\footnotesize   (A) Cell-free protein expression system [adapted with permission from \citep{hershewe2020cell}]. (B) Electron micrograph of the synthetic cell \textit{JCVI-Syn3.0} (precursor of \textit{Syn3A}) [adapted with permission from \citep{hutchison2016design}]. }
\label{fig:fig4} 
\vspace{-2 mm}
\end{figure}

Bottom-up synthesis aims to build cells \textit{de novo} purely from artificially synthesized inanimate biomolecules --- a tantalizing opportunity to catch where matter becomes life. \textit{In vitro} or ``cell free'' systems like PURE \citep{cui2022cell} successfully replicate  isolated cell processes such as translation \citep{shimizu2005protein}, transcription \citep{cui2022cell}, and replication of simple gene circuits [ Fig. \ref{fig:fig4}A]  \citep{fujiwara2013cooperative}.  These process elements are transplanted into vesicles built from liposomes \citep{herianto2022liposome},  coacervates \citep{lin2023coacervate}, preoteinosomes \citep{maffeis2024synthetic} or polymersomes \citep{discher2002cross, kamat2011engineering}. The resulting ``cell mimics'' \citep{xu2016artificial} are used to engineer and optimize RNA synthesis \citep{adamala2013nonenzymatic}, signaling cascades \citep{elani2014vesicle}, and motility \citep{lemiere2015cell}, as well as to generate energy  \citep{berhanu2019artificial} or even synthesize food. In one example, Bodenschatz, Golsetanian and co-workers developed a minimal synthetic beating axenome, paving the way for synthetic cells capable of motility \citep{guido2022synthetic}. \begin{marginnote}[]
	\entry{cell mimics}{cell-like structures built bottom-up from artificially synthesized chemicals to perform specific cell-like functions}
\end{marginnote}
Synthetic vesicle engineering has led to platforms for self-sustaining protocells, as demonstrated by Boekhoven and co-workers \citep{zozulia2024acyl, donau2020active}. The broader hope is that combining the full complement of essential cell processes will ``boot up'' self-replication \citep{rothschild2024building} and, hence, life. In the language of systems biology: elements of the network have been synthesized, and co-locating them in a vesicle should integrate them into a cooperative, self-propagating, living whole cell. Yet so far no cell mimic has autonomously self-replicated beyond a single generation \citep{xu2016artificial}. If synthetic biology's argument is that only pure synthesis of a living cell reveals how matter becomes life \citep{rothschild2024building}, non-living cell mimics are not yet there.  

 The bottom-up approach's key fault line is physical compartmentalization. In living cells, temporary but durable compartments guide biomolecules together to promote desired biochemical reactions, prevent deleterious cross-reactions, and selectively couple pathways \citep{jia2019bottom} ({\em e.g.}, sequestering glycolytic enzymes in G-bodies to promote glycolysis under hypoxia \citep{jin2017glycolytic} or promoting gene transcription by pooling chromatin-associated co-factors \citep{boija2018transcription}). Such coordination is essential for complex biosynthesis \citep{olivi2021towards}. Crucially, bottom-up synthetic cells struggle with cell division, often failing to bring together the replicated genome and compartmentalized cell machinery (ribosomes, proteins, etc.) in each daughter cell. The best success thus far executes a single generation \citep{olivi2021towards}. Overcoming this barrier requires mastering physical organization and physics' role in transforming matter to life. Spatz and coworkers have made progress with this barrier by integrating reverse-engineered  enclosures and other functional modules in eukaryotes, e.g., building biofidelic vesicles capable of onboarding molecules using microfluidic technologies, and controlling onboarded-genome structure via DNA nanotechnologies \citep{weiss2018sequential,  gopfrich2020dynamic, lussier2021ph}.

 In contrast, the top-down approach deconstructs a natural living cell to a minimal self-replicating form. This is done by deleting bits of its genome until the cell can no longer carry out one or more essential life functions \citep{hutchison1999global}.  The J. Craig Venter Institute achieved this, starting with the first full sequencing of \textit{Mycoplasma genitalium}, a bacterium with the smallest naturally-occurring genome \citep{fraser1995minimal}. In a \textit{tour de force} of synthetic biology, their team developed tools to systematically silence genes \citep{hutchison1999global} and identify the minimal genome capable of life \citep{glass2006essential}, then transplanted it into a host that lived, grew, and divided \citep{lartigue2007genome}; and later synthesized the genome \textit{de novo} \citep{gibson2008complete}. They pooled these techniques to construct the minimal cell capable of life with the latest version being \textit{JCVI-Syn3A} at the time of this article {[Fig. \ref{fig:fig4}B]  \citep{hutchison2016design, breuer2019essential}. Calling Syn3A a ``synthetic cell'' is disputed \citep{hutchison2016design} because only the genome is synthetic; the host and its cytoplasm are natural. Nonetheless, it is our opinion that \textit{Syn3A}'s minimal genome makes it a convincing place to peel back complexity and reveal where matter becomes life. 

Overall, top-down and bottom-up approaches provide two fundamental insights about how matter becomes life. It is clear that life requires compartmentalization and a minimal set of genes. 

\section{Compartmentalization and condensation: Hints of \textit{where} matter becomes life}
\label{sec:compartmentalization}

Systems biology's reaction network descriptions are missing a sense of where inside a cell they take place and how that physical organization affects them. Perhaps this contributes to the failure to couple them into whole living synthetic cells. Indeed, the cell itself is defined first and foremost by its enclosure, which sequesters biomolecules, nutrients and ions within a watery compartment that allows diffusive transport throughout the cell. \begin{marginnote}[]
	\entry{homeostasis}{a self-regulating process of maintaining internal environment of the cell to preserve its autonomy}
\end{marginnote}This spatiotemporal coordination of biomolecules maintains cell autonomy (homeostasis) \citep{keener2009cellular}, which favors life processes.  There are also compartments within cells, both membrane-bound and membrane-less. The former includes the nucleus, the Golgi complex, the endoplasmic reticulum, and mitochondria \citep{casares2019membrane, lombard2014once} --- but these are not present in prokaryotes or Archaea, and thus are not essential for life and are unlikely to reveal where matter becomes life.

\begin{textbox}[t]\section{MEMBRANELESS COMPARTMENTS AND COACERVATES}
Intracellular compartments can form without a membrane, enclosed only by liquid surface tension in droplet-like regions \citep{wheeler2018controlling}. The idea that cell autonomy requires compartmentalization, but that such compartments need not include a membrane, was first put forth by Oparin, who speculated that life emerged in a pre-membrane world where ``coagula'' of protein-like colloids compartmentalized biomolecules to favor life processes \citep{oparin1924proiskhozhdenie}. In 1929, Bungenberg-de Jong coined the term ``coacervates'' and, with Kruyt \citep{bungenberg1929coacervation}, extended Oparin's ideas to modern cells, speculating that cytoplasm is a system of coacervates. The idea of cytoplasmic coacervates gained traction in the last two decades, as membrane-less compartments were identified and their importance in cell function recognized. 
\end{textbox}

\begin{figure}
\centering
\includegraphics[width=\linewidth]{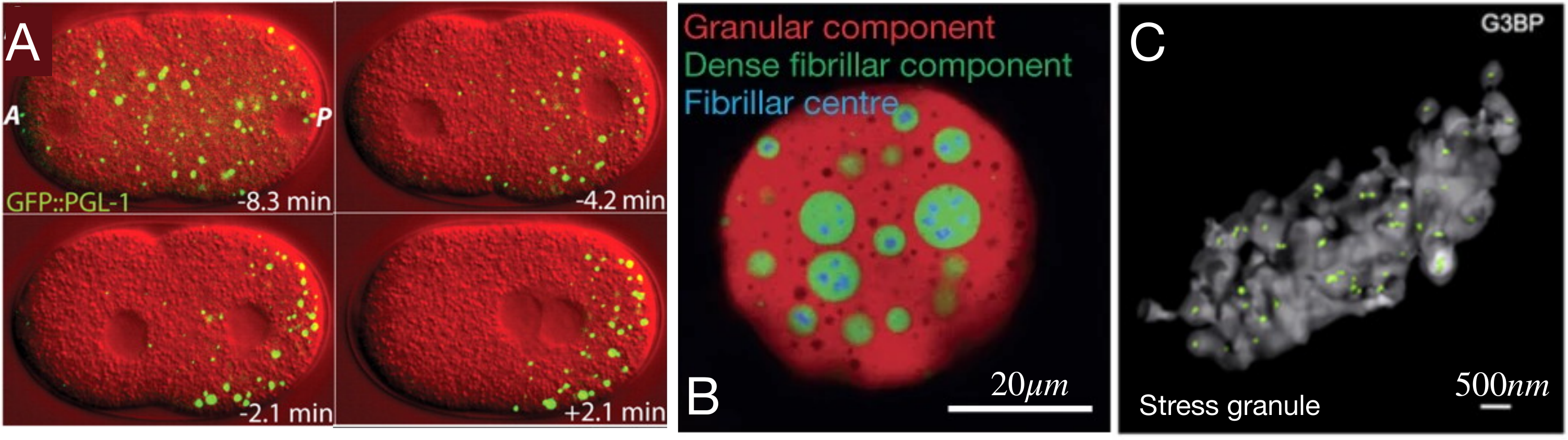}
\vspace{-3 mm}
\caption{\footnotesize  (A) Localization of liquid-like P-granule over time [adapted with permission from \citep{brangwynne2009germline}].  (B) Fluorescent image of nucleoli in \textit{X. laevis} germinal vesicle [adapted with permission from \citep{gouveia2022capillary}]. (C) Image of mRNA-protein stress granule of yeast cells [adapted with permission from \citep{jain2016atpase}].} 
\label{fig:fig5} 
\vspace{-2 mm}
\end{figure}

Membrane-less compartments are biomolecular condensates \citep{wheeler2018controlling}, which form on the fly, emerging and disappearing multiple times throughout the cell cycle. Such compartments can form through liquid-liquid phase separation (LLPS) [Fig. \ref{fig:fig5}] \citep{shin2017liquid}, as demonstrated by Brangwynne and Hyman and co-workers \citep{brangwynne2009germline, wheeler2018controlling}, and J{\"u}licher and co-workers \citep{lee2013spatial, laha2024chemical}. The \textit{in vivo} observation of LLPS in membrane-less P-granules by Hyman and Brangwynne ushered in a novel focus on the cell's physical architecture and the spatial distribution of the cellular constituents \citep{brangwynne2009germline}. There are several recent reviews summarizing progress in this subfield, e.g. from Fare {\em et al.} \cite{fare2021higher}. These dynamic condensates host reaction pathways, regulate competing metabolic pathways, and chaperone cell reorganization during cell division \citep{wheeler2018controlling}. They also facilitate huge gains of function, for example, regulating the biogenesis of the Central Dogma's molecular building blocks.  Membrane-less compartment dysfunction is implicated in diseases such as ALS, Alzheimer's, Parkinson's, and cancer \citep{shin2017liquid}. Recent work suggests that `condensates' contain a range of structures formed by aggregation or gelation, where physical interactions are precisely tied to biological functions \citep{alberti2016aberrant}.  Interestingly, these essential dynamic compartments often evade synthetic biology, hinting at why cell mimics cannot yet fully self-propagate. Yet, condensate formation is easily predicted by common abiogenic phase behavior, e.g. molecular and polymer theories for aggregation and liquid-liquid phase separation (LLPS).  In other words, the matter-to-life transition point or region is still not obvious.

The bacterial nucleoid is a more obvious nexus between physical dynamics and the Central Dogma. The nucleoid hosts gains in function that transform matter to life \citep{crick1958protein}. A membrane-less nucleoid occupies a distinct region of the cell -- well-separated from the cell wall in most prokaryotes, except \textit{Mollicutes} ({\em e.g.}, \hyperref[sec:wcm]{Section 7}. The nucleoid's backbone carries its genome while its interstitial pores host transcription factors and other molecules.  The pore size distribution selectively excludes biomolecules based on size and charge \citep{valverde2024macromolecular, joyeux2018segregative}; many bacterial cells exclude ribosomes from the nucleoid, spatially segregating transcription from translation \citep{valverde2024macromolecular}. \begin{marginnote}[]	
\entry {Nucleoid}{entangled porous network of bacterial DNA, nucleoid- associating proteins, RNA polymerase, and other enzymes.}
\end{marginnote} 
Some nucleoids can regulate cell growth by dynamically remodeling themselves in response to environmental stimuli. For instance, \textit{E. coli}'s nucleoid adapts to thermal stress by compacting itself and, in tandem, altering gene expression.  G{\"o}pfrich and co-workers mimic these effects by dynamically actuating DNA nanostructures inside synthetic cell mimics \citep{gopfrich2020dynamic}.  We speculate that this physico-genetic restructuring is part of a broader expression-regulation cycle that optimizes cell fitness, a fundamental gain of function \cite{valverde2024macromolecular, SRZ-inprep2024}. 

Beyond transcription and its connection to translation, the nucleoid hosts adaption-essential gains in function. But these functions require a cooperative response over the entire cell, a functionality underpinned by transport and flow.

\section{Transport and flow: spatial resolution is key to modeling cells}

Fluid-borne motion undergirds cellular function, including homeostasis and adaptation to environmental conditions. Diffusive (passive) and chemical energy-driven (active) motion transports biomolecules throughout the cell and its compartments, helping them grow, divide, and respond to environmental changes. For example, under thermal stress, a coordinated set of transport and reactions cascade across the membrane and through the cell to prevent protein misfolding. Conformational changes in membrane-bound proteins pump enzymes into the cell, which diffuse through cytoplasm to activate transcription factors. Brownian motion transports them into the nucleoid where they bind to DNA and up-regulate heat shock proteins (HSPs). HSPs migrate out to cytoplasm and prevent protein misfolding \citep{craig1993heat}. This example highlights the range of coordinated transport, from ion pumps to molecular motors, cellular-scale diffusion, macroscopic flow, and condensation. 

Molecular pumps powered by ATP drive much of the cell’s cross-membrane transport. They import nutrients, maintain ion concentration, and expel waste to maintain cell homeostasis \citep{van2001ion} by pumping ions and molecules against an opposing gradient \citep{alberts2002carrier}. Examples include ABC transporters, proton pumps, and ionic pumps like sodium-potassium pumps \citep{alberts2002carrier}, which import glucose  \mbox{\citep{alberts2002carrier}} for ATP synthesis, the cell's energy currency. When cellular glucose is depleted, Na-K pumps eject intracellular sodium ions which bind to and charge extracellular glucose molecules. \begin{marginnote}
	\entry{ATP}{Adenosine triphosphate is a molecule used for energy storage and conversion in cells.}
\end{marginnote}
A nearby charge-sensitive protein channel pulls glucose-laden ions back in the cell. Modeling uncovered the mechanisms of cross-membrane transport, including regulation via hydration forces between phospholipid bilayers \citep{pertsin2007origin} and signal propagation by nutrient stimulus \citep{alim2017mechanism}. Engineering these pumps could sustain nutrient uptake in cell mimics, addressing the self-propagation limit \citep{xu2021transmembrane}.

Molecular motors \citep{sweeney2018motor} in eukaryotes, also powered by ATP, provide deterministic, cell-spanning cargo transport and carry out cytoskeletal reorganization. Mechanistically, these motors attach to cytoskeletal elements and 1) process (`climb') along them to carry attached macromolecular loads; or, 2) exert forces that reshape and reconfigure microtubules to form and position the cell’s mitotic spindle \citep{fink2011external}. The three most prevalent motor proteins in eukaryotes – kinesin, dyenin, and myosin – all carry macromolecular cargo. Only kinesin and dyenin remodel microtubules and reposition mitotic spindles \citep{sweeney2018motor}.  Motor proteins can also reshape the cell’s membrane by transporting vesicles across it \citep{dehmelt2010spatial, schmick2014kras}}. 

\begin{marginnote}
	\entry{Motor proteins}{a/k/a molecular motors include mysoin, which moves along actin filaments; and kinesin and dyenin, which move along tubulin filaments (microtubules).}
	\end{marginnote}
 
\begin{figure}
\centering
\includegraphics[width=\linewidth]{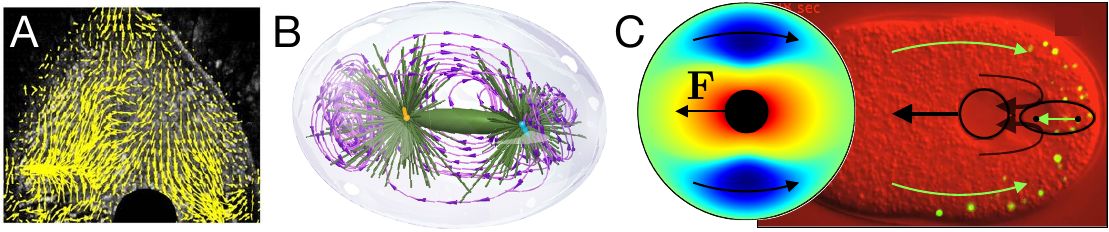}
\vspace{-3 mm}
\caption{\footnotesize  Cellular-scale transport in cells. Cell spanning cytoplasmic streaming in (A)drosophila oocyte observed experimentally [adapted with permission from \citep{ganguly2012cytoplasmic}] and (B) simulation of \textit{C. elegans} embryo  [adapted with permission from \citep{wu2024laser}]. (C) Concentration gradient driven diffusion arising from hydrodynamic coupling predicted by Confined Stokesian dynamics simulations [adapted with permission \citep{aponte2016simulation}] matches patterns in \textit{C. elegans} [adapted with permission from \citep{brangwynne2009germline}]}
\label{fig:fig7} 
\vspace{-4 mm}
\end{figure}

Biomolecular motion drags and displaces fluid, inducing secondary flows like Brownian drift, arising from concentration gradients \citep{sunol2023confined}, as well as cytoplasmic streaming, arising from molecular motors moving along intracellular filaments \citep{nazockdast2017cytoplasmic} (cf Fig. \ref{fig:fig7}). Brownian motion -- driven by thermal energy, $kT$, where $k$ is Boltzmann's constant and $T$ is the absolute solvent temperature -- creates passive transport. Passive self-diffusion mixes cytoplasm, creates osmotic pressure, and drives combinatoric matching of mRNA translation molecules \citep{maheshwari2023colloidal}. Hydro-entropic forces also produce collective and gradient diffusion, a deterministic concentration-gradient-induced motion  \citep{batchelor1976brownian, batchelor1983diffusion, batchelor1977effect, brady1994long, zia2018active}. The coupled diffusive oscillation of \textit{E. coli}’s Min System proteins, which locates the septum for cell division \citep{adler1967miniature}, is an important example. There are excellent reviews of diffusion in general and in cells, {\em e.g.}, \citep{maheshwari2019colloidal, zia2018active, brangwynne2009intracellular}.

In contrast, molecular motors’ active motion drags bulk fluid in cells \citep{sweeney2018motor} inducing streaming flows.  For example, myosin’s motion along filament bundles continuously circulates fluid; this cytoplasmic streaming entrains and rapidly advects biomolecules throughout the cell \citep{goldstein2015physical}. This flow sweeps cargo proteins toward target organelles \citep{verchot2010cytoplasmic} and helps establish polarity during division \citep{nazockdast2017cytoplasmic}. For example, Grill, J{\"u}licher and co-workers showed how interplay of PAR-complex-based feedback with cortical flow establishes cell polarity in \textit{Caenorhabditis elegans} embryos \citep{gross2019guiding}. The hydrodynamics of this cell-spanning spiraling flow were characterized by Goldstein and co-workers \citep{goldstein2015physical}, Shelley and co-workers \citep{nazockdast2017cytoplasmic}, Zia and co-workers \cite{aponte2016simulation}, and Saintillain \citep{saintillan2018extensile} with connections to physiological functions \citep{tominaga2015molecular}. Kamiya and co-workers and Goldstein, among others, provide comprehensive reviews on active transport \mbox{\citep{chowdhury2005physics}}, cytoplasmic streaming \mbox{\citep{kamiya1981physical, verchot2010cytoplasmic}}, and general cell flow \mbox{\citep{freund2012fluid}}. 

\begin{marginnote}[]
	\entry{cytoplasmic streaming}{fluid flow arising from the entrainment of fluid due to the motion of cargo-carrying motor proteins along filament bundles}
\end{marginnote}

Condensation in cells marries these perspectives. As discussed in \hyperref[sec:compartmentalization]{Section 5}, aggregation, phase separation, vitrification, and gelation in cells creates microcompartments that facilitate reactions and create elements of the reaction network modeled in synthetic and systems biology. 

Overall, transport in cells is well understood in two limiting regimes that divide study along two disciplines. Structural and molecular biologists and biochemists have elucidated atomistic and molecular-scale cross-membrane transport and vesicle trafficking. Biophysicists and hydrodynamicists have shown how streaming flows and diffusion orchestrate whole-cell reorganization. This transport is represented to varying degrees in whole-cell models, providing potential connections to where material transport becomes life.

\section{Reuniting biology, physics, and fluid mechanics: whole-cell modeling}
\label{sec:wcm}

Whole-cell modeling aims to predict a cell’s complete life cycle and phenotypes by encompassing functions every of gene and biomolecule \citep{georgouli2023multi}. The task is formidable: solving highly-coupled stochastic partial differential equations that couple motion in three dimensions and time \citep{smith2019spatial} --- for thousands, millions, or billions of particles, depending on the resolution sought --- with reaction chemistry. The associated transport processes are separated by orders of magnitude in length and time scales, further complicating this task \cite{ouaknin2021parallel}. No existing model solves these equations for every atom and molecule in any cell. Simplifications to make whole-cell modeling tractable use a range of coarse-graining paradigms, with recent efforts aiming to connect biology-centric and physics-centric approaches, representing enough chemistry and physics to instantiate the Central Dogma but neglecting obstructive detail. Such cross-disciplinary thinking may yield further insight about where matter becomes life.

Covert's models of \textit{Mycoplasmas} and \textit{E. coli} set the bar for comprehensive whole-cell modeling using constraint-based ordinary differential equations (ODE) systems representing cellular process networks \citep{karr2012whole, karr2015networkpainter, macklin2020simultaneous, ahn2022expanded}. His trailblazing whole-cell model of \textit{Mycoplasma genitalium} integrated 28 modules to kinetically model metabolism, replication, repair, and more \citep{karr2012whole}. Its first key innovation was judiciously smearing-out explicit transport and spatial complexity, reserving computational power for highly-coupled pathways that predict temporal variation in nutrient uptake, protein production, and more. Its second key innovation accounted for how reaction pathways influence each other during whole-cell function. The model's algorithm integrated together a multitude of individual reaction pathways ---  e.g. glycolytic energy generation and amino acid synthesis --- encoding how each influences the other, changing rates throughout the cell's growth cycle \citep{feist2007genome, suthers2010improved}. Covert's models predicted rates for varying growth media, temperature, pH, and gene disruption. Its novel capability to model over multiple cell generations is a gateway to modeling mutations and adaptation.

\begin{marginnote}
	\entry{reaction pathway}{sequence of chemical reactions from reactants to final products, with potential intermediates formed along the way, giving detailed molecular-level mechanisms of the complete reaction.}
	\end{marginnote}

\begin{figure}
\centering
\includegraphics[width=\linewidth]{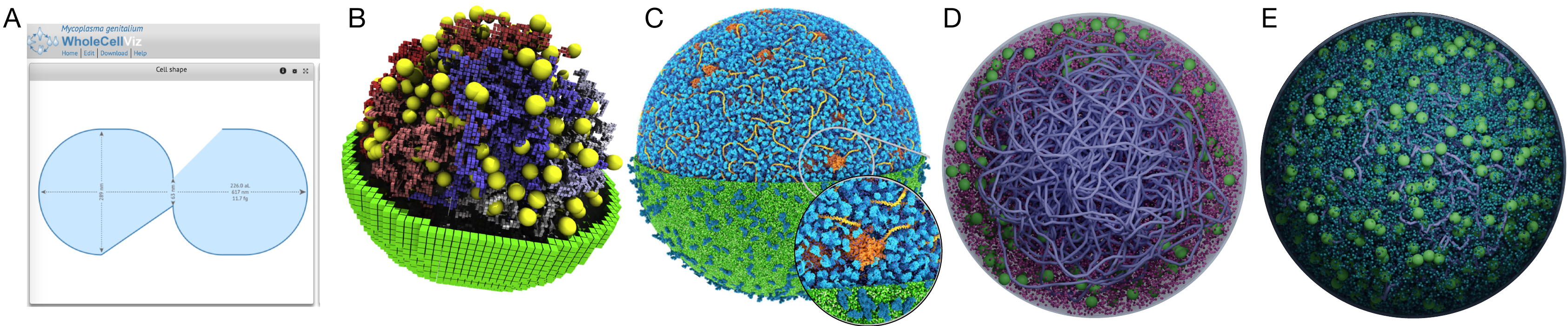}
\vspace{-1 mm}
\caption{\footnotesize  (A) Simulation snapshot of the kinetics-based whole-cell model of \textit{Mycoplasma genitalium} from live simulator at https://www.wholecellviz.org/viz.php [adapted with permission from \citep{karr2012whole}]. (B) Lattice model representation of \textit{JCVI-Syn3A} cell [adapted with permission \citep{thornburg2022fundamental}]. (C) Static Martini-based whole-cell model of \textit{JCVI-Syn3A} [adapted from \citep{stevens2023molecular} (Creative Commons CC BY License)].  Dynamic simulation of spatially-resolved, bio-colloidal whole-cell model of \textit{JCVI-Syn3A} (D) highlighting 100-bp resolution nucleoid and (E) highlighting simply-modeled translation molecules and proteins \citep{GR-inprep2024,roure2024modeling}.}
\label{fig:fig8} 
\vspace{-6 mm}
\end{figure}

In contrast to Covert’s whole-cell model, Luthey-Schulten and coworkers' models simplify the comprehensive partial differential equation to an ODE by coupling individual reaction pathways to a three-dimensional lattice (Figure \ref{fig:fig8}B) \citep{thornburg2022fundamental}. This matrix records the concentrations and positions of proteins, ribosomes, RNAs and DNA as point particles. Spatial resolution is circumvented by permitting reactions to occur only within assigned matrix elements. Lattice data provide initial conditions for the master ODEs for individual reaction pathways \citep{ breuer2019essential} and the network coupling emerges from lattice constraints. The matrix approach is an important step toward representing how molecules' spatial distribution affects reaction rates.  The model provides robust synthesis predictions over a single life cycle. However, the authors acknowledge that the model assumes constant diffusion rates throughout the cell cycle --- approximated for dilute conditions --- while crowding is well known to affect cell rate processes \citep{maheshwari2023colloidal, scott2010interdependence}.

\begin{marginnote}
	\entry{crowded diffusion}{Brownian diffusion is a montonically decreasing function of packing fraction, thus, diffusion slows in more crowded cells.}
\end{marginnote}

Current trends in whole-cell modeling are moving toward explicit physical co-location of reacting biomolecules, eschewing the simplification to kinetics-only ODEs that can miss interactions by smearing out spatial resolution. Several approaches carefully balance biochemical and physical accuracy with computational efficiency.

All-atom whole-cell models attempt to take the farthest leap away from the ODE approach. At first glance, an intriguing 2023 paper by Stevens et al. appeared to achieve the monumental task of dynamically evolving nearly a billion molecules in a whole-cell model of the synthetic cell \textit{JCVI-Syn3A} \citep{stevens2023molecular}. Unfortunately, while the whole-cell images are beautiful, no provided data support the claim that the model can execute dynamic functions (Figure \ref{fig:fig8}C). Nonetheless, the model's static biochemical detail is impressive. The authors systematically incorporated coarse-grained macromolecular representations using Martini, a prolifically successful coarse-graining force-field approach \citep{marrink2007martini, marrink2023two, maritan2022building} that uses Molecular Dynamics (MD) modeling alongside detailed biochemistry to produce individual macromolecules' dynamics over nanoseconds. Martini is quite successful at retaining biochemical information --- for example, base-pair level detail encoded into coarse-grained DNA \citep{marrink2007martini}. \begin{marginnote}
	\entry{Martini model}{coarse-graining (CG) framework that maps detailed atomistic biochemical and physical interactions between a set of molecules onto a CG bead amenable to and more efficiently used in MD simulations. }
\end{marginnote}
But reactions are one of its two limitations. For example, the reactions involved in ribosomal translation of mRNA cannot yet be calculated within the Martini algorithm. This is because the current Martini algorithm requires monitoring every intermediate product and every transition state in a reaction pathway, a formidable computational task in one process, let alone a system of nearly a billion particles. However, efforts to incorporate reactions for biomolecules are underway \citep{sami2023reactive}. The second limitation returns to the physics: solving highly-coupled stochastic partial differential equations in three dimensions and time, i.e., the Langevin equation \citep{lemons1908theorie}. In Stevens \textit{et al}'s Martini approach, the authors perform complete MD simulations, explicitly modeling every solvent (water) molecule and every biomolecule in cytoplasm. To get dynamics over all required time scales, the model needs sufficient computational power to solve the Langevin equation for motion of each molecule. With current global computing power, solving these equations for one full cell cycle would take multiples of years \citep{ouaknin2021parallel}.
	
	Another whole-cell modeling approach invokes a key simplification to improve computational efficiency: bypassing the complexity of modeling individual solvent molecules and ions.  The Brownian dynamics (BD) and Stokesian dynamics (SD) algorithms \citep{BradyBossis88, Brady93b, durlofsky1987dynamic, sierou2001accelerated, banchio2003accelerated, aponte2022confined,aponte2018equilibrium, gonzalez2021impact, ouaknin2021parallel} bypass explicit modeling of every solvent molecule, instead incorporating the mean and fluctuating effects of solvent-mediated interactions. Both approaches, based on the overdamped Langevin equation, calculate particle displacements that occur due to explicitly-imposed stochastic Brownian forces, as well as entropic exclusion. Stokesian dynamics couples solvent-mediated interactions, accounting for an infinitude of reflected interactions between all particles simultaneously, with Accelerated Stokesian Dynamics (ASD) having $O(N\log N)$ efficiency. The most recent enhancements of ASD include a massively parallelized version \citep{ouaknin2021parallel} that can partition a million or more particles across thousands of processes and one that includes size polydispersity, interparticle attraction, and other surface forces \citep{JZ-inprep2024, JHZK-inprep2024}. However, the Fast Fourier Transforms underlying its speed prohibit modeling a finite membrane enclosure. 
\begin{marginnote}
	\entry{size polydispersity}{a system of particles of varying size. This variation permits higher packing and can lead to layered structure. Macroscopic example: a jar packed with marbles can be further packed by adding pebbles, then adding sand.}
\end{marginnote}	Alternatively, Confined Stokesian Dynamics (CSD) rigorously models a spherical enclosure, but its efficiency is hindered by the finite domain calculations \citep{aponte2022confined, aponte2018equilibrium, gonzalez2021impact}. Nonetheless CSD has produced useful insights: Aponte-Rivera and Zia demonstrated that hydrodynamic coupling between biomolecular motion and the cell wall induces gradient-driven motion implicated in P. granules' motion during establishment of polarity in \textit{C. elegans} \citep{aponte2016simulation}, and Gonzalez , Aponte-Rivera and Zia showed that confinement, hydrodynamic interactions, and size polydispersity produce size segregation consistent with cytoplasmic organization \citep{gonzalez2021impact}. However, the complexities of modeling the infinitely-coupled hydrodynamic interactions within an enclosure continue to hinder progress with this approach. We are exploring using AI and machine-learning methods to handle some of the more complicated couplings.
	
	Between the freely-draining BD limit and the strong hydrodynamically-interacting SD limit, there are very effective approaches that incorporate near-field and pair-level far-field hydrodynamic interactions to model biological dynamics in cells. As an example, Saintillan, Shelley, and co-workers have used this approach in an extensive body of work modeling chromatin organization in cellular compartments. \begin{marginnote}
		\entry{freely-draining limit}{hydrodynamic (solvent-mediated) interactions between macromolecules are negligibly weak, so the motion of one does not entrain others. Common situation for charged particles.}
	\end{marginnote}
	Their model includes the forces arising from motor-protein-driven complex interactions, which they show underlies much of the chromatin organization critical for genomic function \citep{saintillan2018extensile, mahajan2022euchromatin, redemann2017c}; they also show how coarse-grained microhydrodynamics can predict  gene expression \citep{mahajan2022euchromatin}. Shelley, Needleman, and co-workers modeled cellular-scale flow-induced organelle and mitotic spindle positioning for cell division \citep{nazockdast2017cytoplasmic, shinar2011model}, leading the field in the detailed modeling of these dynamic cytoplasmic elements. Alternatively, Golestanian and co-workers used mean-field models of active propulsion of colloids in cells \citep{howse2007self, pollack2022competitive}. In some of these approaches, Brownian motion is also included. These methods effectively capture long-range fluid physics and active transport processes, and are among the very few models of eukaryotic cells. But, in terms of whole-cell modeling, detailed biochemistry is missing.

	Returning again to the idea of evolving a complete, whole-cell system with millions of coupled, reacting biomolecules, the approaches reviewed thus far suggest that explicit modeling of how processes spatially coupled is a missing component in understanding life, but accounting for that physics while also incorporating biochemistry is no simple task.

	One recent approach models detailed Brownian physics and reaction chemistry in a subset of cytoplasm. Zia and co-workers modeled translation elongation in \textit{E. coli}, showing that changes in colloidal stoichiometry speed protein synthesis during faster growth \cite{maheshwari2023colloidal}. This work 	\begin{marginnote}
		\entry{coupled biomolecules}{electrostatic and hydrodynamic interactions between particles couple them together via co-location and entrained motion. Biochemical coupling includes affinities, catalysis, or proximity dependence.}
	\end{marginnote}
	took a hybrid approach by gathering a massive data base from experimental cell biology literature describing the physical contents of \textit{E. coli} over a range of growth rates from 0.6 to 3.0 doublings [dbl]/h. From this, they constructed thousands of `translation voxels' --- subsets of cytoplasm outside the nucleoid, populated with ribosomes, ternary complexes, and native proteins. Together this voxel ensemble sampled complete physiological data for translation molecules. Ribosomes, ternary complexes, and protein crowders were represented explicitly and physically, albeit as simple Brownian spheres. Interactions between these particles obeyed a large database of reaction kinetics for cognate and non-cognate codon/anticodon interactions \mbox{\citep{rudorf2014deducing, gromadski2004kinetic}}, and the physical size of the biomolecules explicitly represented crowding in cytoplasm. As a result, a ternary complex's diffusive search emerges directly from a Brownian walk through crowded cytoplasm, and the combinatoric search included physiological matching and mismatching interactions. The authors leveraged LAMMPS' \mbox{\citep{plimpton1995fast}} massively parallelized architecture, an MD simulator with the capability to implicitly model a solvent, to vastly reduce computational expense by invoking the overdamped limit, thereby modeling viscous drag and Brownian motion but neglecting many-body hydrodynamic interactions. We developed tools to track every biomolecule throughout the translation process. The simulation demonstrated that changes in relative translation molecule abundances, combined with \textit{increased} crowding, led to increased per-ribosome productivity. The model’s physical resolution revealed that  the size-polydisperse molecules crowded around ribosomes changed with growth rate to enable faster matching during faster growth rates. Without physical resolution, faster protein synthesis in faster-growing \textit{E. coli} would remain unexplained. Recent refinement of the simple ribosome model to include the L12 stalk showed that physical `pre-loading' of the ribosome is key to elongation rate \citep{Jenpreload2023}. Still, this is only one process in an \textit{in silico} 
		\begin{marginnote}
		\entry{\textit{in silico}}{computational representation of a process or system, intended to compare with {\em in vivo} and {\em in vitro} experiments.}
	\end{marginnote}
	representation of a cell-free system. We have begun development of a simple whole-cell physics-based model of \textit{E. coli}, populating it with a genome, cytoplasm biomolecules, and an implicit solvent: Valverde-Mendez, Sunol, \textit{et al.} studied how biomolecules migrate through the \textit{E. coli}'s nucleoid in simulations and experiments, to study how electrostatics and size exclusion modulate the cell's spatial organization \citep{valverde2024macromolecular}.  Next steps will include hybridizing our LAMMPS approach with the biomolecular coarse-graining of an MD-based physical resolution \citep{PabloJCP2007, sun2021bottom, snodin2015introducing} to avoid the computational juggernaut of modeling solvent molecules, but still leave open options for how to represent chemical reactions.

	But as stated in \S\ref{sec:systemsbio}, traditional whole-cell modeling's completeness, in some sense, provides too much information. In these vast networks, which process, at what scale, in which pathway, does matter transform to life?  We return to the idea of how whole-cell models can reveal the nexus where matter becomes life. It has been said that  ``what is true for \textit{E. coli} is true for an elephant" \citep{morange2010scientific}; but while \textit{E. coli}'s genetic and proteomic complexity reveals much about adaptation and fitness, its genotype adaptability leverages a multitude of functions that are not strictly essential for life. But \textit{JCVI-Syn3A}, with the smallest genome capable of life, possesses only necessary genes to perform life-essential functions \citep{hutchison2016design}. In other words, this minimal cell can be considered the ``hydrogen atom” of cells Morowitz sought \citep{morowitz1984completeness}.Thus, \textit{Syn3A} is an ideal platform for discovering how compartmentalization, pathways, and multi-scale cooperative physics instantiate life.

	To this end, our group is building a physically-resolved, biochemically-representative dynamic model of the \textit{JCVI-Syn3A} whole cell on a Brownian dynamics platform that can evolve cellular functions over minutes [Fig. \ref{fig:fig8}(D,E)]. With the physically-resolved biomolecules, Brownian motion is explicit and parameter-free; biomolecules exert entropic and soft repulsions, short- and long-range electrostatics, obey kinetics of chemical reactions, and are present in physiologically-accurate relative abundances and proteomics. Detailed genomics and transcriptomics data provided by the J. Craig Venter Institute has been coarse-grained to the 100 base-pair-level into a self-assembled nucleoid (smoothed rendering shown in the figure). The model is in its infancy, but its strength is its capability to rigorously simulate physical dynamics and chemical kinetics with high-resolution coarse graining. Our model bypasses all the complexity of solvent molecules, leaving abundant room for more refined biochemistry. The current path forward focuses on further refined coarse-graining from molecular dynamics simulations \citep{snodin2015introducing}, similar in ethos to the refinements presented in the Stevens \textit{et al}. Martini model discussed above, but the added capability to self-assemble both structured and intrinsically-disordered compartments on-the-fly using pipelines that combine PDB and AlphaFold \citep{yang2024in-prep}.
	
	\section{Minimal cells to reveal the nexus between matter and life}
	\label{sec:min_cell}
	Physically- and biochemically-resolved whole-cell models seem to be a key to uncovering of life-instantiating coordination, delineating the boundary at which synthetic materials become life, and enabling the construction of purely synthetic life from chemicals. This inquiry has coalesced around a new frontier, the ‘matter-to-life' transition, as evidenced by training programs such as the Max Planck Matter-to-Life School (https://mattertolife.maxplanckschools.org/matter-to-life-network) and the Alfred P. Sloan Foundation's Matter-to-Life program (https://sloan.org/programs/research/matter-to-life). Several groups have begun to study the matter-to-life idea in whole cells (cf \S\ref{sec:wcm}) using the \textit{JCVI-Syn3A} minimal cell as a platform (cf \S\ref{sec:SynthCells}). But how well does it represent the fundamental transition to life?
	
	\textit{Syn3A} was derived from \textit{Mycoplasma mycoides}, a bacterium with one of the smallest naturally-occurring genomes capable of life and, because Syn3A contains only life-essential genes, it performs only life-essential functions that are conserved across bacterial species \cite{RancatiNatRevGen2018}. For that reason, \textit{Syn3A} is an ideal proving ground to study the nexus of modern life and matter. As with Monod's adage — what is true for \textit{E. coli} is true for the elephant — one can say, what is true for Syn3A is true for all life. However, \textit{Syn3A} has an important physical thumbprint not shared by all bacterial species. Like many \textit{Mycoplasma} species, its ribosomes are uniformly distributed throughout a cell-spanning nucleoid \cite{GilFmolbio2021}, unlike species such as \textit{E. coli}, which confines ribosomes to an annulus surrounding its centralized nucleoid. Understanding these disparate structures takes on further relevance when comparing the two cells. \textit{Mycoplasma}, \textit{Syn3A}'s precursor, underwent a rapid degenerative evolution from \textit{Firmicutes}, in which it lost most of its regulatory genes and its gene-repair machinery -- a loss essential to its genetic minimality \cite{RazMMBE1998}. As a result, \textit{Mycoplasmas} evolved to survive in a specific environment where they can rapidly mutate. In contrast, \textit{E. coli} -- which is on the opposite end of the phylogenetic tree from Mycoplasmas -- evolved to survive a huge range of environmental conditions using a large variety of regulatory and DNA protecting/repair genes. \textit{E. coli} resists mutation and is one of the most adaptive organisms on the planet. We view this combination of phylogenetic separation and vastly different physical organization as making \textit{Syn3A} and \textit{E. coli} ideal foils: one mutates to survive but cannot adapt, while the other is highly adaptive and survives without rapid mutation. Planned survival, a life-essential feature, takes on one of these two disparate genetic states, using one of two very disparate physical states — a difference we hypothesize relates to their different survival strategies. Our preliminary comparative study of these two bacteria suggests that together they are an ideal platform for probing how dynamic remodeling of the nucleoid could reveals the matter/life nexus \citep{SRZ-inprep2024}.

\vspace{-6mm}	

\section{Discussion and outlook}
	The search to understand and engineer life has traced an incredibly rapid trajectory in the last two hundred years, compared to the millennia of philosophical thought that preceded it. Structural biology characterized the building blocks of life and how they process information, while systems biology largely bypassed physical constraints to characterize how  whole cells coordinate molecular interactions to replicate and survive a changing environment. Between these two limits, synthetic biology has harnessed life's machinery to synthesize engineered materials and aims to re-instantiate life by building whole cells \textit{de novo}. But, thus far, progress is capped at cell mimics due in large part to the essential role of physical compartmentalization, physical restructuring, and fluid-borne transport. However, neither localization nor compartmentalization-coupled reaction chemistry is unique to living matter. Somewhere in the physicality of the Central Dogma processes, there are profound gains of function that distinguish matter from life. 

	We speculate an over-arching question going forward is whether bacterial genome's physical dynamics are a key physics-based link missing from the Central Dogma. Fitness points towards this missing link. To live well is to optimize fitness, the capacity to grow at an appropriate pace under an optimal set of growth conditions  \cite{BerkvensJRSI2002}. Fitness requires responding to environmental changes, which bacteria do through gene regulation. Top-down approaches to synthetic biology provide a compelling ``north star" in this search for minimal-genome cells \cite{hutchison2016design}. Our current work focuses on studying \textit{E. coli} and \textit{Syn3A} together because they bookend the adaptation/fitness continuum that is life's hallmark. In particular, we focus on the nucleoid because that is where adaptation takes place. As discussed in \S\ref{sec:compartmentalization}, the bacterial nucleoid is a nexus between physicality and the Central Dogma. It can also be viewed as a compartment in many prokaryotes and hosts much gain in function \citep{crick1958protein} that transforms matter to life. 
	
	Looking forward, we speculate that the nucleoid adaptive physical remodeling --- cascading directly upward to whole-cell function --- hosts the most fundamental matter-to-life transition: the ability to adapt and to instantiate fitness. There is experimental evidence of such remodeling \citep{LawsonCostrbio2004,PeltJmolbio1999}, although its connection to gene regulation and whole-cell transcriptomics has not been recognized. We have begun investigating how this localized, environmentally-induced remodeling over tens of base pairs alters the nucleoid's large-scale structure and cascades up to whole-cell response. We hope that a combined effort between all the approaches reviewed here will lead to insightful whole-cell models that reveal exactly where in the cell's watery milieu that matter becomes life.

\section*{DISCLOSURE STATEMENT}
The authors are not aware of any affiliations, memberships, funding, or financial holdings that might be perceived as affecting the objectivity of this review.

\section*{ACKNOWLEDGMENTS}
This work was supported in part by the Alfred P. Sloan foundation grant number G-2022-19561. The authors greatly acknowledge fruitful collaboration with Dr. Nicole Sharp. We thank Drew Endy, John Glass, Gesse Roure, and Alp M. Sunol for many useful conversations. 

\bibliographystyle{ar-style3}

\begin{thebibliography}{231}
	\expandafter\ifx\csname natexlab\endcsname\relax\def\natexlab#1{#1}\fi
	
	\bibitem{aristotle365anima}
	Aristotle A. 365 BCE.
	\textit{De anima (On the soul)}.
	self-published
	
	\bibitem{aristotle1986anima}
	Aristotle A. 1986.
	\textit{De anima (On the soul). Translated by Edghill}.
	Penguin
	
	\bibitem{mayr1982growth}
	Mayr E. 1982.
	\textit{The growth of biological thought: Diversity, evolution, and
		inheritance}.
	Harvard University Press
	
	\bibitem{AmHtgDict}
	2022.
	{teleology. American Heritage Dictionary of the English Language}.
	\url{https://ahdictionary.com/word/search.html?q=teleology}.
	Accessed: 2024-12-14
	
	\bibitem{schuster2006scientific}
	Schuster JA. 2006.
	The scientific revolution.
	In \textit{Companion to the history of modern science}. Routledge
	
	\bibitem{dobell1932antony}
	Dobell C. 1932.
	\textit{Antony van Leeuwenhoek and his "Little Animals."}.
	Dover Publications
	
	\bibitem{brown1828xxvii}
	Brown R. 1828.
	Xxvii. a brief account of microscopical observations made in the months of
	june, july and august 1827, on the particles contained in the pollen of
	plants; and on the general existence of active molecules in organic and
	inorganic bodies.
	\textit{Phil. Mag.} 4(21):161--173
	
	\bibitem{einstein1906theory}
	Einstein A. 1906.
	{On the theory of the Brownian movement}.
	\textit{Ann. Phys} 19(4):371--381
	
	\bibitem{smoluchowski1906kinetic}
	Smoluchowski M. 1906.
	{The kinetic theory of Brownian molecular motion and suspensions}.
	\textit{Ann. Phys} 21:756--780
	
	\bibitem{perrin1909movement}
	Perrin J. 1909.
	Mouvement brownien et r{\'e}alit{\'e} mol{\'e}culaire.
	\textit{Ann. Chim. Phys.} 18:5--114
	
	\bibitem{haw2006middle}
	Haw M. 2006.
	\textit{Middle World: The Restless Heart of Matter and Life}.
	Springer
	
	\bibitem{schleiden1838beitrage}
	Schleiden M. 1838.
	\textit{Beitr{\"a}ge zur Phytogenesis}.
	Veit et Comp
	
	\bibitem{schwann1839mikroskopische}
	Schwann T. 1839.
	\textit{Mikroskopische Untersuchungen uber die Uebereinstimmung in der Struktur
		und dem Wachsthum der Thiere und Pflanzen}.
	Verlag der Sander'schen Buchhandlung
	
	\bibitem{leewenhoeck1977observation}
	Leewenhoeck Av. 1977.
	Observation, communicated to the publisher by {Mr}. {Antony} van {L}eewenhoeck,
	in a dutch letter of the 9 octob. 1676 here english'd: concerning little
	animals by him observed in rain-well-sea and snow water; as also in water
	wherein pepper had lain infused.
	\textit{Phil. Trans} 12:821--31
	
	\bibitem{bauer1838illustrations}
	Bauer F. 1838.
	\textit{Illustrations of Orchidaceons Plants}.
	James Ridgway and sons
	
	\bibitem{brown1833observations}
	Brown R. 1833.
	\textit{Observations on the organs and mode of fecundation in Orchideae and
		Asclepiadeae}.
	Taylor
	
	\bibitem{mendel1996experiments}
	Mendel G. 1865.
	Experiments in plant hybridization.
	\textit{Verhandlungen des naturforschenden Vereins Br{\"u}nn.}
	
	\bibitem{huxley1942evolution}
	Huxley J. 1942.
	{Evolution: The Modern Synthesis}
	
	\bibitem{darwin2016origin}
	Darwin C. 2016.
	On the origin of species, 1859
	
	\bibitem{de1910intracellular}
	De~Vries H. 1910.
	\textit{Intracellular pangenesis: including a paper on fertilization and
		hybridization}.
	Open Court Publishing Company
	
	\bibitem{muller2010cell}
	M{\"u}ller-Wille S. 2010.
	Cell theory, specificity, and reproduction, 1837--1870.
	\textit{Stud. Hist. Philos. Sci. C} 41(3):225--231
	
	\bibitem{vines1880works}
	Vines SH. 1880.
	The works of carl von n{\"a}geli.
	\textit{Nature} 23(578):78--80
	
	\bibitem{miescher1871ueber}
	Miescher JF. 1871.
	\textit{Ueber die chemische Zusammensetzung der Eiterzellen}
	
	\bibitem{crick1957nucleic}
	Crick FHC. 1957.
	Nucleic acids.
	\textit{Sci. Am.} 197(3):188--203
	
	\bibitem{altmann1889ueber}
	Altmann R. 1889.
	Ueber nucleins{\"a}uren.
	\textit{Arch. Anat. Phys.} 5:524--536
	
	\bibitem{levene1931nucleic}
	Levene PA, Bass LW. 1931.
	\textit{Nucleic acids}.
	Chemical Catalog Company New York
	
	\bibitem{franklin1953molecular}
	Franklin RE, Gosling RG. 1953.
	{Molecular Configuration in Sodium Thymonucleate}.
	\textit{Nature} 171:740--741
	
	\bibitem{watson1953structure}
	Watson JD, Crick F, et~al. 1953.
	A structure for deoxyribose nucleic acid.
	\textit{Nature}
	
	\bibitem{jacob1961genetic}
	Jacob F, Monod J. 1961.
	Genetic regulatory mechanisms in the synthesis of proteins.
	\textit{J. Mol. Biol.} 3(3):318--356
	
	\bibitem{roeder1969multiple}
	Roeder RG, Rutter WJ. 1969.
	{Multiple forms of DNA-dependent RNA polymerase in eukaryotic organisms}.
	\textit{Nature} 224(5216):234--237
	
	\bibitem{ullmann1971cyclic}
	Ullmann A, Contesse G, Crepin M, Gros F, Monod J. 1971.
	\textit{Cyclic AMP and catabolite repression in Escherichia coli}.
	In \textit{The Role of Adenyl Cyclase and Cyclic 3', 5'-AMP in Biological
		Systems: A Colloquium Sponsored by the John E. Fogarty International Center
		for Advanced Study in the Health Sciences, National Institutes of Health,
		November 17-19, 1969, National Institutes of Health, Bethesda, Maryland},
	no.~4, pp.  215. US Government Printing Office
	
	\bibitem{boulikas1992evolutionary}
	Boulikas T. 1992.
	Evolutionary consequences of nonrandom damage and repair of chromatin domains.
	\textit{J. Mol. Evol.} 35:156--180
	
	\bibitem{tomkova2018dna}
	Tomkova M, Schuster-B{\"o}ckler B. 2018.
	{DNA modifications: naturally more error prone?}
	\textit{Trends Genet.} 34(8):627--638
	
	\bibitem{duan2018reduced}
	Duan C, Huan Q, Chen X, Wu S, Carey LB, et~al. 2018.
	{Reduced intrinsic DNA curvature leads to increased mutation rate}.
	\textit{Genome Biol.} 19:1--12
	
	\bibitem{curry2015structural}
	Curry S. 2015.
	Structural biology: a century-long journey into an unseen world.
	\textit{Interdiscip. Sci. Rev.} 40(3):308--328
	
	\bibitem{sahl2017fluorescence}
	Sahl SJ, Hell SW, Jakobs S. 2017.
	Fluorescence nanoscopy in cell biology.
	\textit{Nat. Rev. Mol. Cell Biol.} 18(11):685--701
	
	\bibitem{schermelleh2019super}
	Schermelleh L, Ferrand A, Huser T, Eggeling C, Sauer M, et~al. 2019.
	Super-resolution microscopy demystified.
	\textit{Nat. Cell Biol.} 21(1):72--84
	
	\bibitem{bacic2020recent}
	Bacic L, Sabantsev A, Deindl S. 2020.
	Recent advances in single-molecule fluorescence microscopy render structural
	biology dynamic.
	\textit{Curr. Opin. Struct. Biol.} 65:61--68
	
	\bibitem{burley2023rcsb}
	Burley SK, Bhikadiya C, Bi C, Bittrich S, Chao H, et~al. 2023.
	Rcsb protein data bank (rcsb. org): delivery of experimentally-determined pdb
	structures alongside one million computed structure models of proteins from
	artificial intelligence/machine learning.
	\textit{Nucleic Acids Res.} 51(D1):D488--D508
	
	\bibitem{morimoto1971molecular}
	Morimoto H, Lehmann H, Perutz MF. 1971.
	Molecular pathology of human haemoglobin: stereochemical interpretation of
	abnormal oxygen affinities.
	\textit{Nature} 232(5310):408--413
	
	\bibitem{wang2016exploring}
	Wang J, Luttrell J, Zhang N, Khan S, Shi N, et~al. 2016.
	\textit{Exploring human diseases and biological mechanisms by protein structure
		prediction and modeling}.
	Springer
	
	\bibitem{renaud2020structural}
	Renaud JP. 2020.
	\textit{Structural biology in drug discovery: methods, techniques, and
		practices}.
	John Wiley \& Sons
	
	\bibitem{sacquin2016bridging}
	Sacquin-Mora S. 2016.
	Bridging enzymatic structure function via mechanics: a coarse-grain approach.
	\textit{Methods Enzymol.} 578:227--248
	
	\bibitem{pauling1953proposed}
	Pauling L, Corey RB. 1953.
	A proposed structure for the nucleic acids.
	\textit{Proc. Natl. Acad. Sci. U. S. A.} 39(2):84--97
	
	\bibitem{crick1954complementary}
	Crick FHC, Watson JD. 1954.
	The complementary structure of deoxyribonucleic acid.
	\textit{Proc. R. Soc. Lond. A Math. Phys. Sci.} 223(1152):80--96
	
	\bibitem{crick1958protein}
	Crick FH. 1958.
	\textit{On protein synthesis}.
	In \textit{Symp. Soc. Exp. Biol.}, vol.~12, pp. ~8
	
	\bibitem{thieffry1998forty}
	Thieffry D, Sarkar S. 1998.
	Forty years under the central dogma.
	\textit{Trends Biochem. Sci.} 23(8):312--316
	
	\bibitem{hewitt2020negative}
	Hewitt SM. 2020.
	Negative consequences of the central dogma
	
	\bibitem{pauling1951structure}
	Pauling L, Corey RB, Branson HR. 1951.
	The structure of proteins: two hydrogen-bonded helical configurations of the
	polypeptide chain.
	\textit{Proc. Natl. Acad. Sci. U. S. A.} 37(4):205--211
	
	\bibitem{hodgkin1951crystallography}
	Hodgkin D, Perutz M. 1951.
	Crystallography.
	\textit{Annu. Rep. Prog. Chem.} 48:361--382
	
	\bibitem{whitford2013proteins}
	Whitford D. 2013.
	\textit{Proteins: structure and function}.
	John Wiley \& Sons
	
	\bibitem{hargittai2015linus}
	Hargittai I. 2010.
	Linus pauling’s quest for the structure of proteins.
	\textit{Struct. Chem.} 21:1--7
	
	\bibitem{whitford2015protein}
	Whitford PC, Onuchic JN. 2015.
	What protein folding teaches us about biological function and molecular
	machines.
	\textit{Curr. Opin. Struct. Biol.} 30:57--62
	
	\bibitem{american2005so}
	{American Association for the Advancement of Science and American Association
		for the Advancement of Science and others}. 2005.
	So much more to know.
	\textit{Science} 309(5731):78--102
	
	\bibitem{chiti2006protein}
	Chiti F, Dobson CM. 2006.
	Protein misfolding, functional amyloid, and human disease.
	\textit{Annu. Rev. Biochem.} 75(1):333--366
	
	\bibitem{valastyan2014mechanisms}
	Valastyan JS, Lindquist S. 2014.
	Mechanisms of protein-folding diseases at a glance.
	\textit{Dis. Models Mech.} 7(1):9--14
	
	\bibitem{berman2000protein}
	Berman HM, Westbrook J, Feng Z, Gilliland G, Bhat TN, et~al. 2000.
	The protein data bank.
	\textit{Nucleic Acids Res.} 28(1):235--242
	
	\bibitem{berman2003announcing}
	Berman H, Henrick K, Nakamura H. 2003.
	Announcing the worldwide protein data bank.
	\textit{Nat. Struct. Mol. Biol.} 10(12):980--980
	
	\bibitem{wwpdb2019protein}
	wwPDB consortium. 2019.
	{Protein Data Bank: the single global archive for 3D macromolecular structure
		data}.
	\textit{Nucleic Acids Res.} 47(D1):D520--D528
	
	\bibitem{qu2009guide}
	Qu X, Swanson R, Day R, Tsai J. 2009.
	A guide to template based structure prediction.
	\textit{Curr. Protein Pept. Sci.} 10(3):270--285
	
	\bibitem{fiser2010template}
	Fiser A. 2010.
	Template-based protein structure modeling.
	\textit{Comput. biol.} :73--94
	
	\bibitem{bonneau2001rosetta}
	Bonneau R, Tsai J, Ruczinski I, Chivian D, Rohl C, et~al. 2001.
	Rosetta in casp4: progress in ab initio protein structure prediction.
	\textit{Proteins: Struct., Funct., Bioinf.} 45(S5):119--126
	
	\bibitem{rohl2004protein}
	Rohl CA, Strauss CE, Misura KM, Baker D. 2004.
	Protein structure prediction using rosetta.
	In \textit{Methods Enzymol.}, vol. 383. Elsevier
	
	\bibitem{jumper2021highly}
	Jumper J, Evans R, Pritzel A, Green T, Figurnov M, et~al. 2021.
	Highly accurate protein structure prediction with alphafold.
	\textit{Nature} 596(7873):583--589
	
	\bibitem{evans2021protein}
	Evans R, O’Neill M, Pritzel A, Antropova N, Senior A, et~al. 2021.
	Protein complex prediction with alphafold-multimer.
	\textit{bioRxiv} :2021--10
	
	\bibitem{varadi2022alphafold}
	Varadi M, Anyango S, Deshpande M, Nair S, Natassia C, et~al. 2022.
	Alphafold protein structure database: massively expanding the structural
	coverage of protein-sequence space with high-accuracy models.
	\textit{Nucleic Acids Res.} 50(D1):D439--D444
	
	\bibitem{varadi2024alphafold}
	Varadi M, Bertoni D, Magana P, Paramval U, Pidruchna I, et~al. 2024.
	Alphafold protein structure database in 2024: providing structure coverage for
	over 214 million protein sequences.
	\textit{Nucleic Acids Res.} 52(D1):D368--D375
	
	\bibitem{deiana2019intrinsically}
	Deiana A, Forcelloni S, Porrello A, Giansanti A. 2019.
	Intrinsically disordered proteins and structured proteins with intrinsically
	disordered regions have different functional roles in the cell.
	\textit{PloS one} 14(8):e0217889
	
	\bibitem{rezaei2012intrinsically}
	Rezaei-Ghaleh N, Blackledge M, Zweckstetter M. 2012.
	Intrinsically disordered proteins: from sequence and conformational properties
	toward drug discovery.
	\textit{ChemBioChem} 13(7):930--950
	
	\bibitem{rauscher2015structural}
	Rauscher S, Gapsys V, Gajda MJ, Zweckstetter M, De~Groot BL, Grubm{\"u}ller H.
	2015.
	Structural ensembles of intrinsically disordered proteins depend strongly on
	force field: a comparison to experiment.
	\textit{J. Chem. Theory Comput.} 11(11):5513--5524
	
	\bibitem{strodel2021energy}
	Strodel B. 2021.
	Energy landscapes of protein aggregation and conformation switching in
	intrinsically disordered proteins.
	\textit{J. Mol. Biol.} 433(20):167182
	
	\bibitem{yang2024in-prep}
	Yang TS, Qin J, Zia R, Jarosz DF. 2024.
	A molecular mechanism for the non-amyloid prion transition.
	\textit{In prep}
	
	\bibitem{nussinov2012allosteric}
	Nussinov R, Tsai CJ, Xin F, Radivojac P. 2012.
	Allosteric post-translational modification codes.
	\textit{Trends Biochem. Sci.} 37(10):447--455
	
	\bibitem{macek2019protein}
	Macek B, Forchhammer K, Hardouin J, Weber-Ban E, Grangeasse C, Mijakovic I.
	2019.
	Protein post-translational modifications in bacteria.
	\textit{Nat. Rev. Microbiol.} 17(11):651--664
	
	\bibitem{jin2021multisite}
	Jin F, Gr{\"a}ter F. 2021.
	How multisite phosphorylation impacts the conformations of intrinsically
	disordered proteins.
	\textit{PLoS Comput. Biol.} 17(5):e1008939
	
	\bibitem{kitano2002systems}
	Kitano H. 2002.
	Systems biology: a brief overview.
	\textit{Science} 295(5560):1662--1664
	
	\bibitem{aggarwal2003functional}
	Aggarwal K, Lee HK. 2003.
	Functional genomics and proteomics as a foundation for systems biology.
	\textit{Brief Funct. Genomics} 2(3):175--184
	
	\bibitem{costa2008complex}
	Costa LdF, Rodrigues FA, Cristino AS. 2008.
	Complex networks: the key to systems biology.
	\textit{Genet. Mol. Biol.} 31:591--601
	
	\bibitem{mcadams2000gene}
	McAdams HH, Arkin A. 2000.
	Gene regulation: Towards a circuit engineering discipline.
	\textit{Curr. Biol.} 10(8):R318--R320
	
	\bibitem{walleczek2006self}
	Walleczek J. 2006.
	\textit{Self-organized biological dynamics and nonlinear control: toward
		understanding complexity, chaos and emergent function in living systems}.
	Cambridge Univ. Press
	
	\bibitem{brophy2014principles}
	Brophy JA, Voigt CA. 2014.
	Principles of genetic circuit design.
	\textit{Nature methods} 11(5):508--520
	
	\bibitem{mcadams1998simulation}
	McAdams HH, Arkin A. 1998.
	Simulation of prokaryotic genetic circuits.
	\textit{Annu. Rev. Biophys. Biomol. Struct.} 27(1):199--224
	
	\bibitem{kim2018molecular}
	Kim SG, Noh MH, Lim HG, Jang S, Jang S, et~al. 2018.
	Molecular parts and genetic circuits for metabolic engineering of
	microorganisms.
	\textit{FEMS Microbiol. Lett.} 365(17):fny187
	
	\bibitem{zhao2014mrna}
	Zhao YB, Krishnan J. 2014.
	{mRNA translation and protein synthesis: an analysis of different modelling
		methodologies and a new PBN based approach}.
	\textit{BMC Syst. Biol.} 8:1--24
	
	\bibitem{guillemin2020noise}
	Guillemin A, Stumpf MP. 2020.
	Noise and the molecular processes underlying cell fate decision-making.
	\textit{Phys. Biol.} 18(1):011002
	
	\bibitem{thomas2014phenotypic}
	Thomas P, Popovi{\'c} N, Grima R. 2014.
	Phenotypic switching in gene regulatory networks.
	\textit{Proc. Natl. Acad. Sci. U. S. A.} 111(19):6994--6999
	
	\bibitem{cascante2008metabolomics}
	Cascante M, Marin S. 2008.
	Metabolomics and fluxomics approaches.
	\textit{Essays Biochem.} 45:67--82
	
	\bibitem{kosuri2007tabasco}
	Kosuri S, Kelly JR, Endy D. 2007.
	Tabasco: A single molecule, base-pair resolved gene expression simulator.
	\textit{BMC bioinformatics} 8:1--15
	
	\bibitem{karr2012whole}
	Karr JR, Sanghvi JC, Macklin DN, Gutschow MV, Jacobs JM, et~al. 2012.
	A whole-cell computational model predicts phenotype from genotype.
	\textit{Cell} 150(2):389--401
	
	\bibitem{maritan2022building}
	Maritan M, Autin L, Karr J, Covert MW, Olson AJ, Goodsell DS. 2022.
	Building structural models of a whole mycoplasma cell.
	\textit{J. Mol. Biol.} 434(2):167351
	
	\bibitem{macklin2020simultaneous}
	Macklin DN, Ahn-Horst TA, Choi H, Ruggero NA, Carrera J, et~al. 2020.
	Simultaneous cross-evaluation of heterogeneous e. coli datasets via mechanistic
	simulation.
	\textit{Science} 369(6502):eaav3751
	
	\bibitem{shah2021review}
	Shah HA, Liu J, Yang Z, Feng J. 2021.
	Review of machine learning methods for the prediction and reconstruction of
	metabolic pathways.
	\textit{Front. Mol. Biosci.} 8:634141
	
	\bibitem{presnell2019systems}
	Presnell KV, Alper HS. 2019.
	Systems metabolic engineering meets machine learning: a new era for data-driven
	metabolic engineering.
	\textit{Biotechnol. J.} 14(9):1800416
	
	\bibitem{hillis2012principles}
	Hillis DM, Sadava DE, Hill RE, Price MV. 2014.
	\textit{Principles of life: for the AP Course}.
	{W H Freeman}
	
	\bibitem{deplazes2009synthetic}
	Deplazes A, Huppenbauer M. 2009.
	Synthetic organisms and living machines: Positioning the products of synthetic
	biology at the borderline between living and non-living matter.
	\textit{Syst. Synth. Biol.} 3:55--63
	
	\bibitem{ginsberg2014synthetic}
	Ginsberg AD, Calvert J, Schyfter P, Elfick A, Endy D. 2014.
	\textit{Synthetic aesthetics: investigating synthetic biology's designs on
		nature}.
	MIT press
	
	\bibitem{morowitz1984completeness}
	Morowitz H. 1984.
	The completeness of molecular biology.
	\textit{Isr. J. Med. Sci.} 20(9):750--753
	
	\bibitem{old1981principles}
	Old RW, Primrose SB. 1981.
	\textit{Principles of gene manipulation: an introduction to genetic
		engineering}, vol.~2.
	Univ of California Press
	
	\bibitem{di2019restriction}
	Di~Felice F, Micheli G, Camilloni G. 2019.
	Restriction enzymes and their use in molecular biology: An overview.
	\textit{J. Biosci.} 44(2):38
	
	\bibitem{brown2020gene}
	Brown TA. 2020.
	\textit{{Gene cloning and DNA analysis: an introduction}}.
	John Wiley \& Sons
	
	\bibitem{benner2005synthetic}
	Benner SA, Sismour AM. 2005.
	Synthetic biology.
	\textit{Nat. Rev. Genet.} 6(7):533--543
	
	\bibitem{hanczyc2020engineering}
	Hanczyc MM. 2020.
	Engineering life: A review of synthetic biology.
	\textit{Artif. Life} 26(2):260--273
	
	\bibitem{bashor2018understanding}
	Bashor CJ, Collins JJ. 2018.
	Understanding biological regulation through synthetic biology.
	\textit{Annu. Rev. of Biophys.} 47(1):399--423
	
	\bibitem{ren2020recent}
	Ren J, Lee J, Na D. 2020.
	Recent advances in genetic engineering tools based on synthetic biology.
	\textit{J. Microbiol.} 58:1--10
	
	\bibitem{weninger2016key}
	Weninger A, Killinger M, Vogl T. 2016.
	{Key methods for synthetic biology: genome engineering and DNA assembly}.
	In \textit{Synthetic biology}. Springer
	
	\bibitem{cui2022cell}
	Cui Y, Chen X, Wang Z, Lu Y. 2022.
	Cell-free pure system: evolution and achievements.
	\textit{BioDesign Research} 2022
	
	\bibitem{shimizu2005protein}
	Shimizu Y, Kanamori T, Ueda T. 2005.
	Protein synthesis by pure translation systems.
	\textit{Methods} 36(3):299--304
	
	\bibitem{fujiwara2013cooperative}
	Fujiwara K, Katayama T, Nomura SiM. 2013.
	Cooperative working of bacterial chromosome replication proteins generated by a
	reconstituted protein expression system.
	\textit{Nucleic Acids Res.} 41(14):7176--7183
	
	\bibitem{herianto2022liposome}
	Herianto S, Chien PJ, Ho JaA, Tu HL. 2022.
	Liposome-based artificial cells: From gene expression to reconstitution of
	cellular functions and phenotypes.
	\textit{Biomater. Adv.} 142:213156
	
	\bibitem{lin2023coacervate}
	Lin Z, Beneyton T, Baret JC, Martin N. 2023.
	Coacervate droplets for synthetic cells.
	\textit{Small Methods} 7(12):2300496
	
	\bibitem{maffeis2024synthetic}
	Maffeis V, Heuberger L, Nikoleti{\'c} A, Schoenenberger CA, Palivan CG. 2024.
	Synthetic cells revisited: Artificial cells construction using polymeric
	building blocks.
	\textit{Adv. Sci.} 11(8):2305837
	
	\bibitem{discher2002cross}
	Discher BM, Bermudez H, Hammer DA, Discher DE, Won YY, Bates FS. 2002.
	Cross-linked polymersome membranes: vesicles with broadly adjustable
	properties.
	\textit{J. Phys. Chem. B.} 106(11):2848--2854
	
	\bibitem{kamat2011engineering}
	Kamat NP, Katz JS, Hammer DA. 2011.
	Engineering polymersome protocells.
	\textit{J. Phys. Chem. Lett.} 2(13):1612--1623
	
	\bibitem{xu2016artificial}
	Xu C, Hu S, Chen X. 2016.
	Artificial cells: from basic science to applications.
	\textit{Mater. Today} 19(9):516--532
	
	\bibitem{adamala2013nonenzymatic}
	Adamala K, Szostak JW. 2013.
	{Nonenzymatic template-directed RNA synthesis inside model protocells}.
	\textit{Science} 342(6162):1098--1100
	
	\bibitem{elani2014vesicle}
	Elani Y, Law RV, Ces O. 2014.
	Vesicle-based artificial cells as chemical microreactors with spatially
	segregated reaction pathways.
	\textit{Nat. Commun} 5(1):5305
	
	\bibitem{lemiere2015cell}
	Lemi{\`e}re J, Carvalho K, Sykes C. 2015.
	Cell-sized liposomes that mimic cell motility and the cell cortex.
	In \textit{Methods Cell Biol.}, vol. 128. Elsevier
	
	\bibitem{berhanu2019artificial}
	Berhanu S, Ueda T, Kuruma Y. 2019.
	Artificial photosynthetic cell producing energy for protein synthesis.
	\textit{Nat. Commun} 10(1):1325
	
	\bibitem{guido2022synthetic}
	Guido I, Vilfan A, Ishibashi K, Sakakibara H, Shiraga M, et~al. 2022.
	A synthetic minimal beating axoneme.
	\textit{Small} 18(32):2107854
	
	\bibitem{zozulia2024acyl}
	Zozulia O, Kriebisch CM, Kriebisch BA, Soria-Carrera H, Ryadi KR, et~al. 2024.
	Acyl phosphates as chemically fueled building blocks for self-sustaining
	protocells.
	\textit{Angewandte Chemie} 136(30):e202406094
	
	\bibitem{donau2020active}
	Donau C, Sp{\"a}th F, Sosson M, Kriebisch BA, Schnitter F, et~al. 2020.
	Active coacervate droplets as a model for membraneless organelles and
	protocells.
	\textit{Nat. Commun} 11(1):5167
	
	\bibitem{rothschild2024building}
	Rothschild LJ, Averesch NJ, Strychalski EA, Moser F, Glass JI, et~al. 2024.
	Building synthetic cells --- from the technology infrastructure to cellular
	entities.
	\textit{ACS Synth. Biol.} 13(4):974--997
	
	\bibitem{hershewe2020cell}
	Hershewe J, Kightlinger W, Jewett MC. 2020.
	Cell-free systems for accelerating glycoprotein expression and
	biomanufacturing.
	\textit{J. ind. microbiol. biotech.} 47(11):977--991
	
	\bibitem{hutchison2016design}
	Hutchison~III CA, Chuang RY, Noskov VN, Assad-Garcia N, Deerinck TJ, et~al.
	2016.
	Design and synthesis of a minimal bacterial genome.
	\textit{Science} 351(6280):aad6253
	
	\bibitem{jia2019bottom}
	Jia H, Schwille P. 2019.
	Bottom-up synthetic biology: reconstitution in space and time.
	\textit{Curr. Opin. Biotechnol.} 60:179--187
	
	\bibitem{jin2017glycolytic}
	Jin M, Fuller GG, Han T, Yao Y, Alessi AF, et~al. 2017.
	Glycolytic enzymes coalesce in g bodies under hypoxic stress.
	\textit{Cell Rep.} 20(4):895--908
	
	\bibitem{boija2018transcription}
	Boija A, Klein IA, Sabari BR, Dall’Agnese A, Coffey EL, et~al. 2018.
	Transcription factors activate genes through the phase-separation capacity of
	their activation domains.
	\textit{Cell} 175(7):1842--1855
	
	\bibitem{olivi2021towards}
	Olivi L, Berger M, Creyghton RN, De~Franceschi N, Dekker C, et~al. 2021.
	Towards a synthetic cell cycle.
	\textit{Nat. Commun} 12(1):4531
	
	\bibitem{weiss2018sequential}
	Weiss M, Frohnmayer JP, Benk LT, Haller B, Janiesch JW, et~al. 2018.
	Sequential bottom-up assembly of mechanically stabilized synthetic cells by
	microfluidics.
	\textit{Nat. Mater.} 17(1):89--96
	
	\bibitem{gopfrich2020dynamic}
	G{\"o}pfrich K, Urban MJ, Frey C, Platzman I, Spatz JP, Liu N. 2020.
	{Dynamic actuation of DNA-assembled plasmonic nanostructures in microfluidic
		cell-sized compartments}.
	\textit{Nano Lett.} 20(3):1571--1577
	
	\bibitem{lussier2021ph}
	Lussier F, Schr{\"o}ter M, Diercks NJ, Jahnke K, Weber C, et~al. 2021.
	{pH-triggered assembly of endomembrane multicompartments in synthetic cells}.
	\textit{ACS Synth. Biol.} 11(1):366--382
	
	\bibitem{hutchison1999global}
	Hutchison~III CA, Peterson SN, Gill SR, Cline RT, White O, et~al. 1999.
	Global transposon mutagenesis and a minimal mycoplasma genome.
	\textit{Science} 286(5447):2165--2169
	
	\bibitem{fraser1995minimal}
	Fraser CM, Gocayne JD, White O, Adams MD, Clayton RA, et~al. 1995.
	The minimal gene complement of mycoplasma genitalium.
	\textit{Science} 270(5235):397--404
	
	\bibitem{glass2006essential}
	Glass JI, Assad-Garcia N, Alperovich N, Yooseph S, Lewis MR, et~al. 2006.
	Essential genes of a minimal bacterium.
	\textit{Proc. Natl. Acad. Sci. U. S. A.} 103(2):425--430
	
	\bibitem{lartigue2007genome}
	Lartigue C, Glass JI, Alperovich N, Pieper R, Parmar PP, et~al. 2007.
	Genome transplantation in bacteria: changing one species to another.
	\textit{Science} 317(5838):632--638
	
	\bibitem{gibson2008complete}
	Gibson DG, Benders GA, Andrews-Pfannkoch C, Denisova EA, Baden-Tillson H,
	et~al. 2008.
	Complete chemical synthesis, assembly, and cloning of a mycoplasma genitalium
	genome.
	\textit{Science} 319(5867):1215--1220
	
	\bibitem{breuer2019essential}
	Breuer M, Earnest TM, Merryman C, Wise KS, Sun L, et~al. 2019.
	Essential metabolism for a minimal cell.
	\textit{Elife} 8:e36842
	
	\bibitem{keener2009cellular}
	Keener J, Sneyd J. 2009.
	Cellular homeostasis.
	In \textit{Mathematical Physiology: I: Cellular Physiology}. Springer
	
	\bibitem{casares2019membrane}
	Casares D, Escrib{\'a} PV, Rossell{\'o} CA. 2019.
	Membrane lipid composition: effect on membrane and organelle structure,
	function and compartmentalization and therapeutic avenues.
	\textit{Int. J. Mol. Sci.} 20(9):2167
	
	\bibitem{lombard2014once}
	Lombard J. 2014.
	Once upon a time the cell membranes: 175 years of cell boundary research.
	\textit{Biol. Direct} 9:1--35
	
	\bibitem{wheeler2018controlling}
	Wheeler RJ, Hyman AA. 2018.
	Controlling compartmentalization by non-membrane-bound organelles.
	\textit{Philos. Trans. R. Soc. B Biol. Sci.} 373(1747):20170193
	
	\bibitem{oparin1924proiskhozhdenie}
	Oparin A. 1924.
	Proiskhozhdenie zhizny (the origin of life, ann. synge. trans.) in.
	\textit{The origin of life}
	
	\bibitem{bungenberg1929coacervation}
	Bungenberg~de Jong H, Kruyt H. 1929.
	\textit{Coacervation (partial miscibility in colloid systems)}.
	In \textit{Proc. K. Ned. Akad. Wet}, vol.~32, pp.  849--856
	
	\bibitem{brangwynne2009germline}
	Brangwynne CP, Eckmann CR, Courson DS, Rybarska A, Hoege C, et~al.
	2009{\natexlab{a}}.
	Germline p granules are liquid droplets that localize by controlled
	dissolution/condensation.
	\textit{Science} 324(5935):1729--1732
	
	\bibitem{gouveia2022capillary}
	Gouveia B, Kim Y, Shaevitz JW, Petry S, Stone HA, Brangwynne CP. 2022.
	Capillary forces generated by biomolecular condensates.
	\textit{Nature} 609(7926):255--264
	
	\bibitem{jain2016atpase}
	Jain S, Wheeler JR, Walters RW, Agrawal A, Barsic A, Parker R. 2016.
	Atpase-modulated stress granules contain a diverse proteome and substructure.
	\textit{Cell} 164(3):487--498
	
	\bibitem{shin2017liquid}
	Shin Y, Brangwynne CP. 2017.
	Liquid phase condensation in cell physiology and disease.
	\textit{Science} 357(6357):eaaf4382
	
	\bibitem{lee2013spatial}
	Lee CF, Brangwynne CP, Gharakhani J, Hyman AA, J{\"u}licher F. 2013.
	Spatial organization of the cell cytoplasm by position-dependent phase
	separation.
	\textit{Phys. Rev. Lett.} 111(8):088101
	
	\bibitem{laha2024chemical}
	Laha S, Bauermann J, J{\"u}licher F, Michaels TC, Weber CA. 2024.
	Chemical reactions regulated by phase-separated condensates.
	\textit{arXiv preprint arXiv:2403.05228}
	
	\bibitem{fare2021higher}
	Fare CM, Villani A, Drake LE, Shorter J. 2021.
	Higher-order organization of biomolecular condensates.
	\textit{Open Biol.} 11(6):210137
	
	\bibitem{alberti2016aberrant}
	Alberti S, Hyman AA. 2016.
	Are aberrant phase transitions a driver of cellular aging?
	\textit{BioEssays} 38(10):959--968
	
	\bibitem{valverde2024macromolecular}
	Valverde-Mendez D, Sunol AM, Bratton BP, Delarue M, Hofmann JL, et~al. 2024.
	Macromolecular interactions and geometrical confinement determine the 3d
	diffusion of ribosome-sized particles in live escherichia coli cells.
	\textit{bioRxiv}
	
	\bibitem{joyeux2018segregative}
	Joyeux M. 2018.
	A segregative phase separation scenario of the formation of the bacterial
	nucleoid.
	\textit{Soft Matter} 14(36):7368--7381
	
	\bibitem{SRZ-inprep2024}
	Sivasankar VS, Roure G, Zia RN. 2024.
	{Colloidal-scale physics-based computational study of bacterial genome
		remodeling and connections to cell fitness}.
	\textit{In prep}
	
	\bibitem{craig1993heat}
	Craig EA, Gambill BD, Nelson RJ. 1993.
	Heat shock proteins: molecular chaperones of protein biogenesis.
	\textit{Microbiol. Rev.} 57(2):402--414
	
	\bibitem{van2001ion}
	van Vliet AH, Bereswill S, Kusters JG. 2001.
	\textit{Ion metabolism and transport}.
	Wiley Online Library
	
	\bibitem{alberts2002carrier}
	Alberts B, Johnson A, Lewis J, Raff M, Roberts K, Walter P. 2002.
	Carrier proteins and active membrane transport.
	In \textit{Mol. Biol. Cell. 4th edition}. Garland Science
	
	\bibitem{pertsin2007origin}
	Pertsin A, Platonov D, Grunze M. 2007.
	Origin of short-range repulsion between hydrated phospholipid bilayers: A
	computer simulation study.
	\textit{Langmuir} 23(3):1388--1393
	
	\bibitem{alim2017mechanism}
	Alim K, Andrew N, Pringle A, Brenner MP. 2017.
	Mechanism of signal propagation in physarum polycephalum.
	\textit{Proc. Natl. Acad. Sci. U. S. A.} 114(20):5136--5141
	
	\bibitem{xu2021transmembrane}
	Xu Z, Hueckel T, Irvine WT, Sacanna S. 2021.
	Transmembrane transport in inorganic colloidal cell-mimics.
	\textit{Nature} 597(7875):220--224
	
	\bibitem{sweeney2018motor}
	Sweeney HL, Holzbaur EL. 2018.
	Motor proteins.
	\textit{Cold Spring Harb Perspect Biol.} 10(5):a021931
	
	\bibitem{fink2011external}
	Fink J, Carpi N, Betz T, B{\'e}tard A, Chebah M, et~al. 2011.
	External forces control mitotic spindle positioning.
	\textit{Nat. Cell Biol.} 13(7):771--778
	
	\bibitem{dehmelt2010spatial}
	Dehmelt L, Bastiaens PI. 2010.
	Spatial organization of intracellular communication: insights from imaging.
	\textit{Nat. Rev. Mol. Cell Biol.} 11(6):440--452
	
	\bibitem{schmick2014kras}
	Schmick M, Vartak N, Papke B, Kovacevic M, Truxius DC, et~al. 2014.
	Kras localizes to the plasma membrane by spatial cycles of solubilization,
	trapping and vesicular transport.
	\textit{Cell} 157(2):459--471
	
	\bibitem{sunol2023confined}
	Sunol AM, Zia RN. 2023.
	Confined brownian suspensions: Equilibrium diffusion, thermodynamics, and
	rheology.
	\textit{J. Rheol.} 67(2):433--460
	
	\bibitem{nazockdast2017cytoplasmic}
	Nazockdast E, Rahimian A, Needleman D, Shelley M. 2017.
	Cytoplasmic flows as signatures for the mechanics of mitotic positioning.
	\textit{Mol. Biol. Cell} 28(23):3261--3270
	
	\bibitem{maheshwari2023colloidal}
	Maheshwari AJ, Sunol AM, Gonzalez E, Endy D, Zia RN. 2023.
	Colloidal physics modeling reveals how per-ribosome productivity increases with
	growth rate in escherichia coli.
	\textit{Mbio} 14(1):e02865--22
	
	\bibitem{batchelor1976brownian}
	Batchelor G. 1976.
	Brownian diffusion of particles with hydrodynamic interaction.
	\textit{J. Fluid Mech.} 74(1):1--29
	
	\bibitem{batchelor1983diffusion}
	Batchelor G. 1983.
	Diffusion in a dilute polydisperse system of interacting spheres.
	\textit{J. Fluid Mech.} 131:155--175
	
	\bibitem{batchelor1977effect}
	Batchelor GK. 1977.
	The effect of brownian motion on the bulk stress in a suspension of spherical
	particles.
	\textit{J. Fluid Mech.} 83(1):97--117
	
	\bibitem{brady1994long}
	Brady JF. 1994.
	The long-time self-diffusivity in concentrated colloidal dispersions.
	\textit{J. Fluid Mech.} 272:109--134
	
	\bibitem{zia2018active}
	Zia RN. 2018.
	Active and passive microrheology: Theory and simulation.
	\textit{Annu. Rev. Fluid Mech.} 50(1):371--405
	
	\bibitem{adler1967miniature}
	Adler H, Fisher W, Cohen A, Hardigree AA. 1967.
	Miniature escherichia coli cells deficient in DNA.
	\textit{Proc. Natl. Acad. Sci. U. S. A.} 57(2):321--326
	
	\bibitem{maheshwari2019colloidal}
	Maheshwari AJ, Sunol AM, Gonzalez E, Endy D, Zia RN. 2019.
	Colloidal hydrodynamics of biological cells: A frontier spanning two fields.
	\textit{Phys. Rev. Fluids} 4(11):110506
	
	\bibitem{brangwynne2009intracellular}
	Brangwynne CP, Koenderink GH, MacKintosh FC, Weitz DA. 2009{\natexlab{b}}.
	Intracellular transport by active diffusion.
	\textit{Trends Cell Biol.} 19(9):423--427
	
	\bibitem{ganguly2012cytoplasmic}
	Ganguly S, Williams LS, Palacios IM, Goldstein RE. 2012.
	Cytoplasmic streaming in drosophila oocytes varies with kinesin activity and
	correlates with the microtubule cytoskeleton architecture.
	\textit{Proc. Natl. Acad. Sci. U. S. A.} 109(38):15109--15114
	
	\bibitem{wu2024laser}
	Wu HY, Kabacao{\u{g}}lu G, Nazockdast E, Chang HC, Shelley MJ, Needleman DJ.
	2024.
	Laser ablation and fluid flows reveal the mechanism behind spindle and
	centrosome positioning.
	\textit{Nat. Physics} 20(1):157--168
	
	\bibitem{aponte2016simulation}
	Aponte-Rivera C, Zia RN. 2016.
	Simulation of hydrodynamically interacting particles confined by a spherical
	cavity.
	\textit{Phys. Rev. Fluids} 1(2):023301
	
	\bibitem{goldstein2015physical}
	Goldstein RE, Van De~Meent JW. 2015.
	A physical perspective on cytoplasmic streaming.
	\textit{Interface focus} 5(4):20150030
	
	\bibitem{verchot2010cytoplasmic}
	Verchot-Lubicz J, Goldstein RE. 2010.
	Cytoplasmic streaming enables the distribution of molecules and vesicles in
	large plant cells.
	\textit{Protoplasma} 240:99--107
	
	\bibitem{gross2019guiding}
	Gross P, Kumar KV, Goehring NW, Bois JS, Hoege C, et~al. 2019.
	Guiding self-organized pattern formation in cell polarity establishment.
	\textit{Nat. Physics} 15(3):293--300
	
	\bibitem{saintillan2018extensile}
	Saintillan D, Shelley MJ, Zidovska A. 2018.
	Extensile motor activity drives coherent motions in a model of interphase
	chromatin.
	\textit{Proc. Natl. Acad. Sci. U. S. A.} 115(45):11442--11447
	
	\bibitem{tominaga2015molecular}
	Tominaga M, Ito K. 2015.
	The molecular mechanism and physiological role of cytoplasmic streaming.
	\textit{Curr. Opin. Plant Biol.} 27:104--110
	
	\bibitem{chowdhury2005physics}
	Chowdhury D, Schadschneider A, Nishinari K. 2005.
	Physics of transport and traffic phenomena in biology: from molecular motors
	and cells to organisms.
	\textit{Phys. Life Rev.} 2(4):318--352
	
	\bibitem{kamiya1981physical}
	Kamiya N. 1981.
	Physical and chemical basis of cytoplasmic streaming.
	\textit{Annu. Rev. Plant Physiol.} 32(1):205--236
	
	\bibitem{freund2012fluid}
	Freund JB, Goetz JG, Hill KL, Vermot J. 2012.
	Fluid flows and forces in development: functions, features and biophysical
	principles.
	\textit{Development} 139(7):1229--1245
	
	\bibitem{georgouli2023multi}
	Georgouli K, Yeom JS, Blake RC, Navid A. 2023.
	Multi-scale models of whole cells: progress and challenges.
	\textit{Front. Cell Dev. Biol.} 11:1260507
	
	\bibitem{smith2019spatial}
	Smith S, Grima R. 2019.
	Spatial stochastic intracellular kinetics: A review of modelling approaches.
	\textit{Bull. Math. Biol.} 81:2960--3009
	
	\bibitem{ouaknin2021parallel}
	Ouaknin GY, Su Y, Zia RN. 2021.
	Parallel accelerated stokesian dynamics with brownian motion.
	\textit{J. Comput. Phys.} 442:110447
	
	\bibitem{karr2015networkpainter}
	Karr JR, Guturu H, Chen EY, Blair SL, Irish JM, et~al. 2015.
	Networkpainter: dynamic intracellular pathway animation in cytobank.
	\textit{BMC bioinformatics} 16:1--7
	
	\bibitem{ahn2022expanded}
	Ahn-Horst TA, Mille LS, Sun G, Morrison JH, Covert MW. 2022.
	An expanded whole-cell model of e. coli links cellular physiology with
	mechanisms of growth rate control.
	\textit{NPJ Syst. Biol. Appl.} 8(1):30
	
	\bibitem{feist2007genome}
	Feist AM, Henry CS, Reed JL, Krummenacker M, Joyce AR, et~al. 2007.
	A genome-scale metabolic reconstruction for escherichia coli k-12 mg1655 that
	accounts for 1260 orfs and thermodynamic information.
	\textit{Mol. Syst. Biol.} 3(1):121
	
	\bibitem{suthers2010improved}
	Suthers PF, Chang YJ, Maranas CD. 2010.
	Improved computational performance of mfa using elementary metabolite units and
	flux coupling.
	\textit{Metab. Eng.} 12(2):123--128
	
	\bibitem{thornburg2022fundamental}
	Thornburg ZR, Bianchi DM, Brier TA, Gilbert BR, Earnest TM, et~al. 2022.
	Fundamental behaviors emerge from simulations of a living minimal cell.
	\textit{Cell} 185(2):345--360
	
	\bibitem{stevens2023molecular}
	Stevens JA, Gr{\"u}newald F, van Tilburg PM, K{\"o}nig M, Gilbert BR, et~al.
	2023.
	Molecular dynamics simulation of an entire cell.
	\textit{Front. Chem.} 11:1106495
	
	\bibitem{GR-inprep2024}
	Roure G, Sivasankar VS, Zia RN. 2024{\natexlab{a}}.
	{Influence of DNA-bending proteins and electrostatics on nucleoid and
		cytoplasmic organization in a minimal bacterial cell}.
	\textit{In prep}
	
	\bibitem{roure2024modeling}
	Roure G, Sivasankar VS, Zia R. 2024{\natexlab{b}}.
	Modeling the effects of hu proteomic interactions in the genome configuration
	of a minimal cell.
	\textit{Bulletin of the American Physical Society}
	
	\bibitem{scott2010interdependence}
	Scott M, Gunderson CW, Mateescu EM, Zhang Z, Hwa T. 2010.
	Interdependence of cell growth and gene expression: origins and consequences.
	\textit{Science} 330(6007):1099--1102
	
	\bibitem{marrink2007martini}
	Marrink SJ, Risselada HJ, Yefimov S, Tieleman DP, De~Vries AH. 2007.
	The martini force field: coarse grained model for biomolecular simulations.
	\textit{J. Phys. Chem. B.} 111(27):7812--7824
	
	\bibitem{marrink2023two}
	Marrink SJ, Monticelli L, Melo MN, Alessandri R, Tieleman DP, Souza PC. 2023.
	Two decades of martini: Better beads, broader scope.
	\textit{Wiley Interdiscip. Rev. Comput. Mol. Sci.} 13(1):e1620
	
	\bibitem{sami2023reactive}
	Sami S, Marrink SJ. 2023.
	Reactive martini: Chemical reactions in coarse-grained molecular dynamics
	simulations.
	\textit{J. Chem. Theory Comput.} 19(13):4040--4046
	
	\bibitem{lemons1908theorie}
	Lemons D, Gythiel A, Langevin P. 1908.
	{Sur la th{\'e}orie du mouvement brownien [On the theory of Brownian motion]}.
	\textit{CR Acad. Sci.(Paris)} 146:530--533
	
	\bibitem{BradyBossis88}
	Brady JF, Bossis G, et~al. 1988.
	Stokesian dynamics.
	\textit{Annu. Rev. Fluid Mech.} 20(1):111--157
	
	\bibitem{Brady93b}
	Brady JF. 1993.
	The rheological behavior of concentrated colloidal dispersions.
	\textit{J.~Chem Phys.} 99:567--581
	
	\bibitem{durlofsky1987dynamic}
	Durlofsky L, Brady JF, Bossis G. 1987.
	Dynamic simulation of hydrodynamically interacting particles.
	\textit{J. Fluid Mech.} 180:21--49
	
	\bibitem{sierou2001accelerated}
	Sierou A, Brady JF. 2001.
	Accelerated stokesian dynamics simulations.
	\textit{J. Fluid Mech.} 448:115--146
	
	\bibitem{banchio2003accelerated}
	Banchio AJ, Brady JF. 2003.
	Accelerated stokesian dynamics: Brownian motion.
	\textit{J. Chem. Phys.} 118(22):10323--10332
	
	\bibitem{aponte2022confined}
	Aponte-Rivera C, Zia RN. 2022.
	The confined generalized stokes-einstein relation and its consequence on
	intracellular two-point microrheology.
	\textit{J. Colloid. Interface Sci.} 609:423--433
	
	\bibitem{aponte2018equilibrium}
	Aponte-Rivera C, Su Y, Zia RN. 2018.
	Equilibrium structure and diffusion in concentrated hydrodynamically
	interacting suspensions confined by a spherical cavity.
	\textit{J. Fluid Mech.} 836:413--450
	
	\bibitem{gonzalez2021impact}
	Gonzalez E, Aponte-Rivera C, Zia RN. 2021.
	Impact of polydispersity and confinement on diffusion in hydrodynamically
	interacting colloidal suspensions.
	\textit{J. Fluid Mech.} 925:A35
	
	\bibitem{JZ-inprep2024}
	Jeong JH, Zia RN. 2024.
	{Influence of polydispersity and hydrodynamics on yield in colloidal gels}.
	\textit{In prep}
	
	\bibitem{JHZK-inprep2024}
	Jeong JH, Zia RN, Koch D, Hormozi S. 2024.
	{A numerical framework to link colloidal physics to the rheology of thermal
		amorphous materials.}
	\textit{In prep}
	
	\bibitem{mahajan2022euchromatin}
	Mahajan A, Yan W, Zidovska A, Saintillan D, Shelley MJ. 2022.
	Euchromatin activity enhances segregation and compaction of heterochromatin in
	the cell nucleus.
	\textit{Phys. Rev. X} 12(4):041033
	
	\bibitem{redemann2017c}
	Redemann S, Baumgart J, Lindow N, Shelley M, Nazockdast E, et~al. 2017.
	C. elegans chromosomes connect to centrosomes by anchoring into the spindle
	network.
	\textit{Nat. Commun} 8(1):1--13
	
	\bibitem{shinar2011model}
	Shinar T, Mana M, Piano F, Shelley MJ. 2011.
	A model of cytoplasmically driven microtubule-based motion in the single-celled
	caenorhabditis elegans embryo.
	\textit{Proc. Natl. Acad. Sci. U. S. A.} 108(26):10508--10513
	
	\bibitem{howse2007self}
	Howse JR, Jones RA, Ryan AJ, Gough T, Vafabakhsh R, Golestanian R. 2007.
	Self-motile colloidal particles: from directed propulsion to random walk.
	\textit{Phys. Rev. Lett.} 99(4):048102
	
	\bibitem{pollack2022competitive}
	Pollack YG, Bittihn P, Golestanian R. 2022.
	A competitive advantage through fast dead matter elimination in confined
	cellular aggregates.
	\textit{New J. Phys.} 24(7):073003
	
	\bibitem{rudorf2014deducing}
	Rudorf S, Thommen M, Rodnina MV, Lipowsky R. 2014.
	Deducing the kinetics of protein synthesis in vivo from the transition rates
	measured in vitro.
	\textit{PLoS Comput. Biol.} 10(10):e1003909
	
	\bibitem{gromadski2004kinetic}
	Gromadski KB, Rodnina MV. 2004.
	{Kinetic determinants of high-fidelity tRNA discrimination on the ribosome}.
	\textit{Mol. Cell} 13(2):191--200
	
	\bibitem{plimpton1995fast}
	Plimpton S. 1995.
	Fast parallel algorithms for short-range molecular dynamics.
	\textit{J. Comput. Phys.} 117(1):1--19
	
	\bibitem{Jenpreload2023}
	Hofmann JL, Yang TS, Sunol AM, Zia RN. 2024.
	{Pre-loading of translation molecules onto ribosomes speeds transport and
		protein synthesis in \textit{Escherichia coli}}.
	\textit{bioRxiv}
	
	\bibitem{PabloJCP2007}
	Knotts TA, Rathore N, Schwartz DC, De~Pablo JJ. 2007.
	{A coarse grain model for DNA}.
	\textit{J. Chem. Phys.} 126(8)
	
	\bibitem{sun2021bottom}
	Sun T, Minhas V, Korolev N, Mirzoev A, Lyubartsev AP, Nordenski{\"o}ld L. 2021.
	Bottom-up coarse-grained modeling of DNA.
	\textit{Front. Mol. Biosci.} 8:645527
	
	\bibitem{snodin2015introducing}
	Snodin BE, Randisi F, Mosayebi M, {\v{S}}ulc P, Schreck JS, et~al. 2015.
	Introducing improved structural properties and salt dependence into a
	coarse-grained model of DNA.
	\textit{J. Chem. Phys.} 142(23)
	
	\bibitem{morange2010scientific}
	Morange M. 2010.
	The scientific legacy of jacques monod.
	\textit{Res. Microbiol.} 161(2):77--81
	
	\bibitem{RancatiNatRevGen2018}
	Rancati G, Moffat J, Typas A, Pavelka N. 2018.
	{Emerging and evolving concepts in gene essentiality}.
	\textit{Nat. Rev. Genet.} 19(1):34--49
	
	\bibitem{GilFmolbio2021}
	Gilbert BR, Thornburg ZR, Lam V, Rashid FZM, Glass JI, et~al. 2021.
	{Generating chromosome geometries in a minimal cell from cryo-electron
		tomograms and chromosome conformation capture maps}.
	\textit{Front. Mol. Biosci.} 8:644133
	
	\bibitem{RazMMBE1998}
	Razin S, Yogev D, Naot Y. 1998.
	{Molecular biology and pathogenicity of mycoplasmas}.
	\textit{Microbiol. Mol. Biol. Rev} 62(4):1094--1156
	
	\bibitem{BerkvensJRSI2002}
	Berkvens A, Chauhan P, Bruggeman FJ. 2022.
	{Integrative biology of persister cell formation: molecular circuitry,
		phenotypic diversification and fitness effects}.
	\textit{J. R. Soc. Interface} 19(194):20220129
	
	\bibitem{LawsonCostrbio2004}
	Lawson CL, Swigon D, Murakami KS, Darst SA, Berman HM, Ebright RH. 2004.
	{Catabolite activator protein: DNA binding and transcription activation}.
	\textit{Curr. Opin. Struct. Biol.} 14(1):10--20
	
	\bibitem{PeltJmolbio1999}
	Pelton JG, Kustu S, Wemmer DE. 1999.
	{Solution structure of the DNA-binding domain of NtrC with three alanine
		substitutions}.
	\textit{J. Mol. Biol.} 292(5):1095--1110
	
\end{thebibliography}

\end{document}